\documentclass[a4paper,fleqn,usenatbib]{mnras}

\usepackage{newtxtext,newtxmath}
\usepackage{lineno}

\usepackage[T1]{fontenc}
\usepackage{ae,aecompl}

\usepackage{graphicx}	
\usepackage{amsmath}	
\usepackage{amssymb}	
\usepackage{multirow}
\usepackage{xcolor}
\usepackage{subfigure}
\usepackage{comment}

\title[Variability of blazars 3C 66A and B2 1633+38]{Quasi-periodic behaviour in the optical and $\gamma$-ray light curves of blazars 3C 66A and B2 1633+38}

\author[J. Otero-Santos et al.]{
J. Otero-Santos,$^{1,2}$\thanks{E-mail: joteros@iac.es (JOS)}
J. A. Acosta-Pulido,$^{1,2}$
J. Becerra Gonz\'alez,$^{2,1}$
C. M. Raiteri,$^{3}$
\newauthor
V. M. Larionov,$^{4,5}$
P. Pe\~nil,$^{6}$
P. S. Smith,$^{7}$ 
C. Ballester Niebla,$^{8}$
G. A. Borman,$^{9}$
\newauthor
M. I. Carnerero,$^{3}$
N. Castro Segura,$^{2,10}$
T. S. Grishina,$^{4}$
E. N. Kopatskaya,$^{4}$
\newauthor
E. G. Larionova,$^{4}$
D. A. Morozova,$^{4}$
A. A. Nikiforova,$^{4,5}$
S. S. Savchenko,$^{4}$
\newauthor
Yu. V. Troitskaya,$^{4}$
I. S. Troitsky,$^{4}$
A. A. Vasilyev$^{4}$ and
M. Villata$^{3}$
\\
$^{1}$Instituto de Astrof\'isica de Canarias (IAC), E-38200 La Laguna, Tenerife, Spain\\
$^{2}$Universidad de La Laguna (ULL), Departamento de Astrof\'isica, E-38206 La Laguna, Tenerife, Spain\\
$^{3}$INAF-Osservatorio Astrofisico di Torino, via Osservatorio 20, 10025 Pino Torinese, Italy\\
$^{4}$Astronomical Institute, St. Petersburg State University, Universitetskij Pr. 28, Petrodvorets, St. Petersburg 198504, Russia\\
$^{5}$Pulkovo Astronomical Observatory of RAS, Pulkovskoye shosse 60, St. Petersburg 196149, Russia\\
$^{6}$IPARCOS and Department of EMFTEL, Universidad Complutense de Madrid, E-28040 Madrid, Spain \\
$^{7}$Steward Observatory, University of Arizona, Tucson, AZ 85721, USA\\
$^{8}$Universidad de La Laguna (ULL), Facultad de Ciencias, E-38206 La Laguna, Tenerife, Spain\\
$^{9}$Crimean Astrophysical Observatory, P/O Nauchny, Crimea, 298409, Russia\\
$^{10}$School of Physics \& Astronomy, University of Southampton, Highfield, Southampton SO17 1BJ, UK
}

\date{Accepted XXX. Received YYY; in original form ZZZ}

\pubyear{2019}
 
\begin{document}
\label{firstpage}
\pagerange{\pageref{firstpage}--\pageref{lastpage}}
\maketitle

\begin{abstract}
We report on quasi-periodic variability found in two blazars included in the Steward Observatory Blazar Monitoring data sample: the BL Lac object 3C 66A and the Flat Spectrum Radio Quasar B2 1633+38. We collect optical photometric and polarimetric data in V and R bands of these sources from different observatories: St. Petersburg University, Crimean Astrophysical Observatory, WEBT-GASP, Catalina Real-Time Transient Survey, Steward Observatory, STELLA Robotic Observatory and Katzman Automatic Imaging Telescope. In addition, an analysis of the $\gamma$-ray light curves from \textit{Fermi}-LAT is included. Three methods are used to search for any periodic behaviour in the data: the Z-transform Discrete Correlation Function, the Lomb-Scargle periodogram and the Weighted Wavelet Z-transform. 
We find evidences of possible quasi-periodic variability in the optical photometric data of both sources with periods of $\sim$3 years for 3C 66A and $\sim$1.9 years for B2 1633+38, with significances between 3$\sigma$ and 5$\sigma$. Only B2 1633+38 shows evidence of this behaviour in the optical polarized data set at a confidence level of 2$\sigma$-4$\sigma$. This is the first reported evidence of quasi-periodic behaviour in the optical light curve of B2 1633+38. 
Also a hint of quasi-periodic behaviour is found in the $\gamma$-ray light curve of B2 1633+38 with a confidence level $\geqslant$2$\sigma$, while no periodicity is observed for 3C 66A in this energy range.
We propose different jet emission models that could explain the quasi-periodic variability and the differences found between these two sources.
\end{abstract}




\section{Introduction}\label{intro}

Blazars are a subclass of Active Galactic Nuclei (AGNs) whose jets closely point to the Earth, characterized by strong flux variability at different timescales and high radio and optical polarization. These objects emit throughout the electromagnetic spectrum, from radio to $\gamma$-rays, showing a Spectral Energy Distribution (SED) with a characteristic double bump structure \citep[see e.g.][]{romero2017}. Blazars can be categorized in two different classes \citep{urry1995}: BL Lac objects and Flat Spectrum Radio Quasars (FSRQs). The differences between them rely on their optical spectra, (almost) featureless in the case of BL Lacs and with an emission-line equivalent width of $EW<5$ \AA \ \citep{bottcher19}, the SED peak positions and the Compton dominance \citep[luminosity relation between the first and the second peak of the SED, see e.g.][]{finke2013}.

Blazars can show variability on a wide range of timescales \citep[see e.g.][]{fan2018, gupta2018}: Intraday variability refers to changes on the order of minutes or hours. Short-term variability involves changes on timescales of days to weeks. Finally, long-term variability concerns variations over timescales of months to years. Various studies have searched for periodicity patterns on large timescales in blazars as e.g. \cite{hovatta2007}, \cite{sandrinelli2016}, \cite{zhang2017a}, \cite{zhang2017b}, \cite{fan2018}. The detection of periodicity in blazars can have important implications for blazar emission models. Periodic behaviour can be interpreted within scenarios such as the presence of a supermassive binary black hole \citep[SBBH, see e.g.][]{lehto1996, Villata99}, periodic changes in the jet emission mechanism like an helical jet \citep[][]{lainela1999, liu2010}, helical structures, recurrent shock front formation and/or disk instabilities \citep{Camenzind92, Marscher85, Tchekhovskoy11}. Thus, the study of periodicity is fundamental to understand the processes taking place in the jets and accretion disks of black holes in active galaxies.

In this work we analyse two sources, 3C 66A (also known as 3FGL J0222.6+4301) and B2 1633+38 (also known as 3FGL J1635.2+3809, 4C 38.41). These two objects were selected among a list of 27 sources observed by the Steward observatory\footnote{\label{noteSteward}\url{http://james.as.arizona.edu/~psmith/Fermi/}} program in support of the \textit{Fermi} $\gamma$-ray observations (from years 2008 to 2018). Twelve targets having more than 25 optical measurements per year were initially surveyed for periodicity. This preliminary examination for periodicity used the Discrete Correlation Function technique \citep{dcf} and 3C 66A and B2 1633+38 were selected since they exhibit a quasi-sinusoidal shape in their auto-correlation curve. Additional data from other observatories were gathered for these blazars, increasing the sampling to more than 200 observations per year. Note that objects showing evidence of periodic variability on longer timescales, such as $\sim$12 yrs for OJ 287 \citep{lehto1996} will be missed in this search.

3C 66A is a BL Lac type object classified as an intermediate synchrotron peaked blazar, according to its SED \citep{ackermann2015}. The redshift of this object is commonly referenced as z=0.444 \citep{miller1978}. Nevertheless, this value has been strongly questioned in several works 
\citep{finke2008, furniss2013, Paiano17, torres-zafra2018}, since its spectrum does not show any emission line. \citet{furniss2013} and \citet{torres-zafra2018} estimated a lower limit of the redshift of z$\geqslant$0.33. 3C 66A has been the target of multiple monitoring campaigns and surveys, showing variability throughout the electromagnetic spectrum on multiple timescales. It was detected by \textit{Fermi}-LAT in high energy (HE, E>100 MeV) $\gamma$-rays \citep{ackermann2016} and by VERITAS and the MAGIC telescopes in the very high energy regime (VHE, E>100 GeV) \citep{acciari2009, aleksic2011}. In the optical V and R bands, a period of 65 days was detected by \cite{lainela1999}, and a period of 156 days was also detected by \cite{fan2018}, while different long-term analyses claimed periods of $\sim$2.5 years \citep{belokon2003, kaur2017} and $\sim$2 years \citep{fan2018}. Infrared intraday variability was also detected by \cite{dediego1997}, with significant spectral index variations associated. A variability on timescales of hours was observed in X-rays by \cite{ghosh1995}. The $\gamma$-ray variability properties were analysed by \cite{vovk2013} with \textit{Fermi}-LAT data. They found variations down to timescales of $\sim$1.5 days. Intraday variability in radio wavelengths has also been studied by \cite{liu2017}, finding only marginal variations.

The other selected source, B2 1633+38, is a low synchrotron peaked source classified as a FSRQ, located at redshift z=1.814 \citep{paris2017}. Strong flux variations were detected at optical wavelengths by \cite{barbieri1977} in the B band. This was confirmed by \cite{villata1997} with the analysis of intraday and short-term variability of the object in the R band. It has been observed throughout the spectrum \citep[see e.g.][]{raiteri2012}, finding hints of possible periods in radio wavelengths of $\sim$1.52 years \citep{hovatta2007} and $\sim$3.07 years \citep{fan2006}. Moreover, \cite{vovk2013} also found variations in the $\gamma$-ray flux of B2 1633+38 down to timescales of $\sim$17 hours. 

In this paper we study the long-term variability of these two blazars. We have large temporal coverage in the optical band, with light curves that span more than 9 years in the different bands. This is an appropriate baseline for long-term periodicity studies and the identification of quasi-periodic oscillations on the order of a few years. This paper is structured as follows: in Section \ref{data_reduction} the data sample and the reduction techniques are introduced, in Section \ref{per_methods} the different methods used in the periodicity analysis are described, and in Sections \ref{results} and \ref{discussion} we present and discuss the final results of the analysis for each source and the possible implications of these results. Finally in Section \ref{conclusions} we end with a summary of the main results and provide possible interpretations.

\section{Observations and data reduction}\label{data_reduction}

We have analysed optical and high energy $\gamma$-ray \textit{Fermi}-LAT data. Typically, $\gamma$-ray emission has been found correlated with the optical emission, so it is expected that periodicity found in the optical band may also show up in $\gamma$-rays \citep{Bonning09, Chatterjee12, Liao14, Carnerero15, Rajput19}.
Both optical photometric and polarimetric data were collected.

\subsection{Optical Data}\label{opt_data}
\subsubsection{Photometry}
The long-term optical data used in this analysis have been obtained from several observatories and surveys, listed in Table \ref{data_table}. The GLAST-AGILE Support Program (GASP) of the Whole Earth Blazar Telescope\footnote{\url{http://www.oato.inaf.it/blazars/webt/}} \citep[WEBT,][]{webt,villata2009} provided V and R-band data for the two sources, which were taken during multiwavelength monitoring campaigns of 3C 66A in 2003--2004 \citep{bottcher2005} and 2007--2008 \citep{bottcher2009,abdo2011}, and of B2 1633+38 in 2008--2011 \citep{raiteri2012}. Additional R-band data on 3C 66A come from the LX-200 0.4m telescope from the St. Petersburg University and the AZT-8 0.7m telescope from the Crimean Astrophysical Observatory \citep[SPB and CrAO,][]{larionov2013}. We also used the rich V and R data sets produced by the Steward Observatory program in support of the {\it Fermi} $\gamma$-ray observations\textsuperscript{\ref{noteSteward}} \citep{paul_smith}, and by the Catalina Real-Time Transient Survey\footnote{\url{http://crts.caltech.edu/}} \citep[CRTS,][]{catalina_survey} in V band. Finally, R-band data obtained by the Katzman Automatic Imaging Telescope\footnote{\url{http://w.astro.berkeley.edu/bait/kait.html}} \citep[KAIT,][]{kait} and by the STELLar Activity Robotic Observatory\footnote{\url{https://www.aip.de/en/research/facilities/stella/}} \citep[STELLA,][]{stella} were utilized. Retrieved data are already flux calibrated, with no need for further data reduction.

\begin{figure*}
	\includegraphics[width=\textwidth]{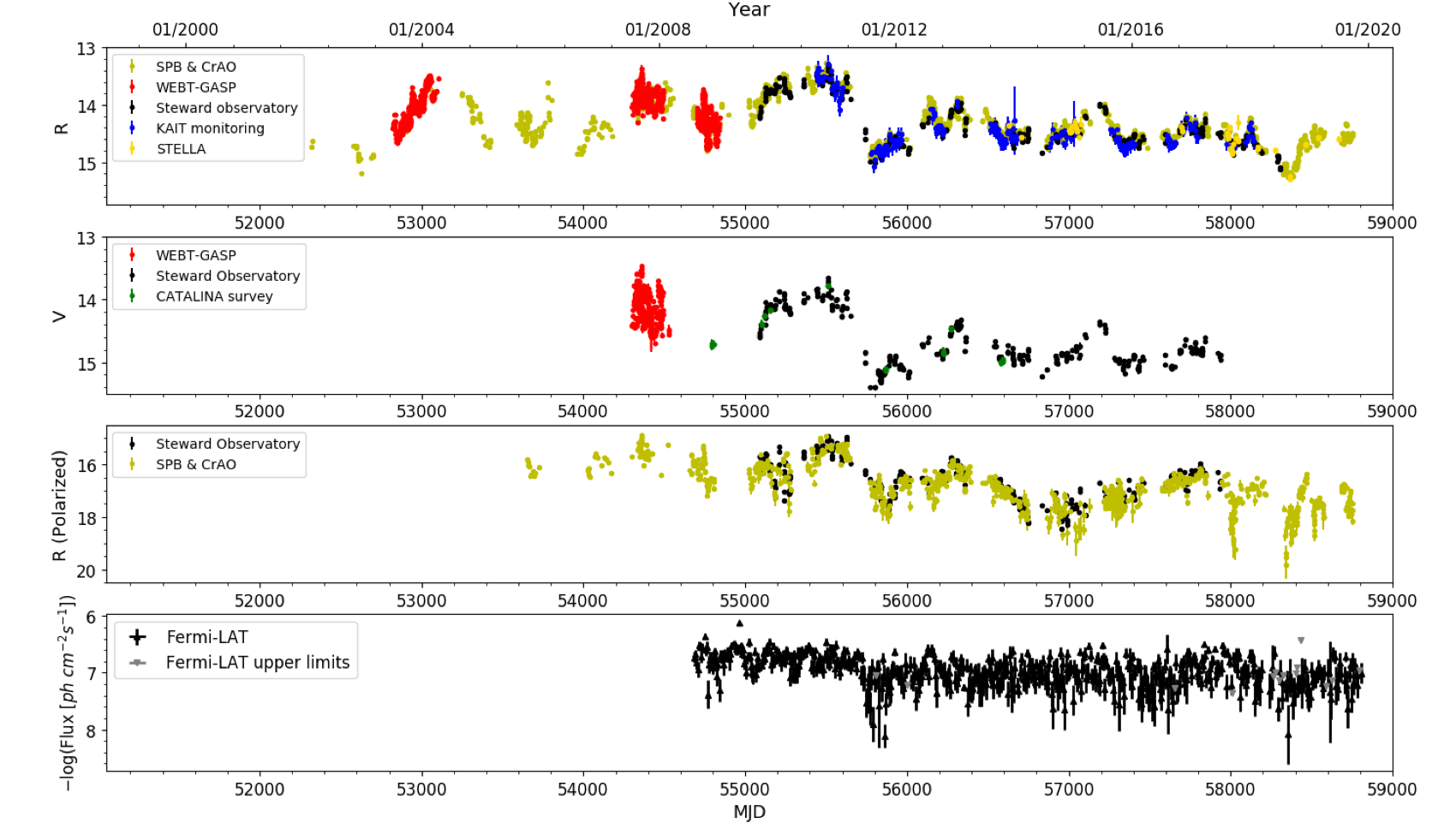}
    \caption{Long-term light curves of 3C 66A. From top to bottom: Optical R band, Optical V band, Polarized magnitude in the optical R band and \textit{Fermi}-LAT $\gamma$-ray light curves. Different colors and symbols denote data from the different observatories used in the analysis.} 
    \label{3c66aLC}
\end{figure*}

\begin{figure*}
	\includegraphics[width=\textwidth]{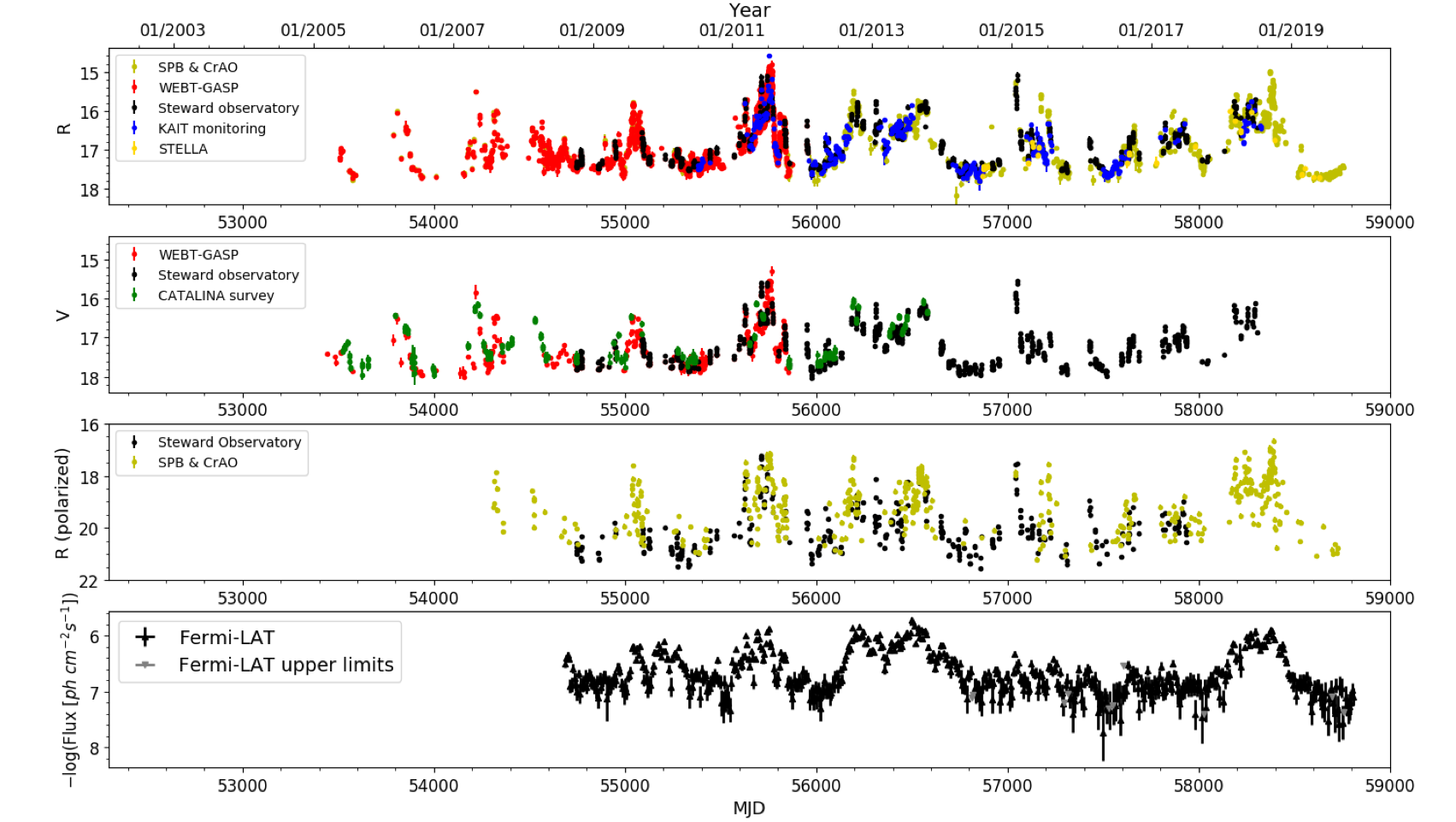}
    \caption{Long-term light curves of B2 1633+38. From top to bottom: Optical R band, Optical V band, Polarized magnitude in the optical R band and \textit{Fermi}-LAT $\gamma$-ray light curves. Different colors and symbols denote data from the different observatories used in the analysis.} 
    \label{b2LC}
\end{figure*}

The R-band light curve of 3C 66A spans about 17.7 years (MJD 52323--58788, 2002 February 17--2019 October 31), while the V-band data cover an interval of $\sim$9.6 years (MJD 54301--58306, 2007 July 20--2018 July 07). Both light curves are shown in Fig. \ref{3c66aLC}. Due to big gaps and limited coverage of the R and V-band light curves of 3C 66A, the data before MJD 52400 and MJD 55000 respectively were not included in the periodicity analysis. The light curves of B2 1633+38 cover about 14.5 years for the R band (MJD 53445--58757, 2005 March 15--2019 October 01) and 12.3 years for the V band (MJD 53444--58306, 2005 March 15--2018 July 07). The resulting light curves are presented in Fig.~\ref{b2LC}. All the optical data were binned on a daily basis (taking the median value of all the values of each night) to perform the analysis in order to eliminate the errors and variations introduced by intraday variability. The calibration between different observatories was checked. All the calibrations were found to be consistent within uncertainties.

\begin{table*}
\scriptsize
\centering
\caption{Long-term optical data sample for 3C 66A and B2 1633+38 from different observatories.}
\label{data_table}
\begin{tabular}{ccccc}
\hline
Source  &  Database &  Band & Monitoring epoch (UT)  & Monitoring epoch (MJD) \\ \cline{1-5} 
\multirow{7}{*}{3C 66A}  & St. Petersburg University \& CrAO &  R   &   2002 Feb. 17 -- 2019 Oct. 31    &  52323--58788  \\ \cline{2-5} 
 &  WEBT-GASP    &    R   & 2003 July 04 -- 2009 Jan. 08   &   52824--54840    \\ \cline{2-5} 
&  WEBT-GASP     &   V   &  2007 July 20 -- 2008 Mar. 03  &   54301--54529   \\ \cline{2-5} 
&  Catalina Real-Time Transient Survey (CRTS)     &  V & 2008 Nov. 23 -- 2013 Oct. 27  &   54803--56592      \\ \cline{2-5} 
&  Steward Observatory      &   R, V   & 2009 Sept. 14 -- 2018 July 07  &  55089--58306    \\ \cline{2-5} 
&  Katzman Automatic Imaging Telescope (KAIT) monitoring     &   R   & 2010 Sept. 03 -- 2018 Jan. 24 &  55443--58142       \\ \cline{2-5}
&  STELLar Activity Robotic Observatory (STELLA)     &   R   &  2013 Dec. 03 -- 2019 July 06  &   56630--58670   \\  \cline{1-5}
\multirow{7}{*}{B2 1633+38}  & WEBT-GASP   &    R   & 2005 Mar. 15 -- 2012 Jan. 30  &   53445--55956    \\ \cline{2-5}
&   WEBT-GASP     &   V   &  2005 Mar. 15 -- 2011 Oct. 22  &    53444--55856   \\ \cline{2-5} 
&  Catalina Real-Time Transient Survey (CRTS)     &    V   & 2005 May 06 -- 2013 Oct. 15  &   53557--56580     \\ \cline{2-5}
& St. Petersburg University \& CrAO &  R   &   2005 July 09 -- 2019 Oct. 01    &  53560--58757  \\ \cline{2-5}
&  Steward Observatory        &    R, V   & 2008 Oct. 06 -- 2018 July 07   &   54745--58306   \\ \cline{2-5} 
&  Katzman Automatic Imaging Telescope (KAIT) monitoring         &    R  &   2010 July 03 -- 2018 July 29   & 55380--58328     \\ \cline{2-5} 
&  STELLar Activity Robotic Observatory (STELLA)     &   R   &  2014 July 31 -- 2019 May 30  &   56869--58634   \\  \cline{1-5}
\end{tabular}
\end{table*}

\subsubsection{Polarimetry}
We have also analysed the variability of the optical polarization in the R band. The data for both sources were retrieved from the ground-based Steward observatory \textit{Fermi} support database and the polarimetric observations of the St. Petersburg University made in St. Petersburg and the Crimean Astrophysical Observatory. 
The Steward R band polarized data are obtained converting the total magnitude into polarized magnitude using the mean polarization degree in the 5000-7000 \AA \ passband. The typical 1$\sigma$ errors for the polarization degree are $<$1\% for the Steward Observatory data and $<$3\% for the St. Petersburg and CrAO data. The light curves corresponding to the polarimetric data are shown in Fig. \ref{3c66aLC} and Fig. \ref{b2LC} for 3C 66A and B2 1633+38, respectively. The polarized light curve of 3C 66A spans about 14 years (MJD 53656--58756, 2005 October 13--2019 September 29). The polarized light curve of B2 1633+38 covers $\sim$12 years (MJD 54313--58729, 2007 August 01--2019 September 03). The polarimetric data were also binned on a daily basis to avoid effects due to intraday variability. The polarization degree has a relative variation with respect to the median of each night typically lower than a 10\% and 12\% for 3C 66A and B2 1633+38, respectively. \cite{hagenthorn2019} found a scatter larger than the observational uncertainties in the rapid variation on the polarization of B2 1633+38. However, the change on the Stokes parameters shows roughly linear relations with the flux variations during the events. This permits to define an average polarization state for each event.

\subsection{High Energy $\gamma$-rays: \textit{Fermi}-LAT Data}\label{fermi_data}

The pair-conversion Large Area Telescope (LAT) on board the \emph{\textit{Fermi}} satellite monitors the $\gamma$-ray sky in survey mode every three hours in the energy range from 20 MeV to $>300$~GeV \citep{Fermitelescope}. For this work, a region of interest (ROI) of 15$^\circ$ was selected centered around the two sources of interest, 3C~66A and B2~1633+382. The data sample included more than 10 years of observations collected by \textit{Fermi}-LAT, from 2008 August 04 to 2019 November 22 (MJD 54682-58809). The data reduction of the events of the Pass8 \texttt{P8R3 source} class was performed with the FermiTools software package version 1.0.10 in the energy range 0.1-300 GeV. To reduce contamination from the limb of the Earth, a zenith angle cut of 90$^\circ$ was applied to the data. The likelihood fit of the data was performed using the recommended Galactic diffuse emission model and isotropic component recommended for the corresponding event selection used in this work\footnote{\url{https://fermi.gsfc.nasa.gov/ssc/data/access/lat/BackgroundModels.html}}. 

A first binned likelihood analysis was performed for the total integration range analyzed within this work. The normalization of both diffuse components in the source model were allowed to freely vary during the spectral fitting. In addition to the source of interest, all the sources included in the 4FGL catalog \citep{fermi4fgl} within a distance of 25$^\circ$ from the sources of interests were included. In the model, up to 15$^\circ$s distance, the targets spectral shape and normalization were left as free parameters, while they were fixed to their 4FGL catalog values for sources located between 15$^\circ$ and 25$^\circ$. From this first fit all the sources located at a distance larger than 15$^\circ$ from the target of interest and with Test Statistic (TS) $<$2 were removed from the model. These simplified models were used for the calculation of the weekly-binned light curves using an unbinned likelihood analysis for this short integration analysis. The resulting light curves are shown in Fig.~\ref{3c66aLC} and Fig.~\ref{b2LC}.

\section{Periodicity Analysis Methods}\label{per_methods}
Three different methods are used in this work to perform the periodicity analysis: the Discrete Correlation Function (DCF), the Lomb-Scargle Periodogram (LS) and the Weighted Wavelet Z-Transform (WWZ). These methods were chosen due to the their different approach when analyzing the data. The DCF is optimized to find repeated patterns along the light curve despite the nature of the variation (e.g. flare-like events or sinusoidal variations). The LS periodogram decomposes the time-series in sinusoidal components, and periodic signals are identified as peaks in the power spectral density (PSD). The WWZ works similarly as the LS, providing information about the periods of the signal and the time associated to those periods. The DCF works in time domain, while the other two work in the Fourier transformed domain. 

\subsection{Discrete Correlation Function}\label{secZDCF}
The DCF is commonly used in the study of AGNs and periodicity searches. While Fourier-based methods of unevenly-sampled data can lead to interpolation and spurious peaks, along with another effects, the DCF is a robust method without interpolation in the temporal domain \citep{dcf}. In this work, an improved DCF is used, named Z-Transformed Discrete Correlation Function (ZDCF), with several changes that improve the performance of the DCF \citep{zdcf}. The main advantages are the use of the z-transform and the equal population binning rather than equal lag $\Delta \tau$. These changes lead to a more robust method to compute the DCF under realistically unfavorable conditions \citep[unevenly sampled and red noise dominated data,][]{zdcf2, zdcf}. This method is limited by the number of points in the light curve. Due to the equal population binning, each bin has to contain at least 11 points, the minimum for a meaningful statistic interpretation. The binning algorithm leads to typical binning of our data of 20-30 days, except for the $\gamma$-ray light curves. Since these curves are almost evenly sampled, the width of the bins in these cases is the same as the sampling interval (7 days).

To test the periodicity hypothesis using the ZDCF method, we study the autocorrelation function. We follow the procedure described in \cite{keplerrotation} to estimate the period of the source. The first step is the identification of the peaks in the autocorrelation curve. To do so, since we are searching for long-term periods, we perform a smoothing of the curve to eliminate the variations and fluctuations introduced by short-term variability. This smoothing is implemented with the LOWESS (LOcally WEighted Scatter-plot Smoother) method \citep{cleveland1979}, useful in detecting trends in noisy data. The basic approach of this method is to start with a local polynomial fit of order \textit{k} (in this work, a linear regression is used) in the neighborhood of each point, and use robust methods to obtain the final curve. A robust measurement of the period is obtained by calculating the median of the time lags that are approximately multiples of the same period.

We perform an estimation of the statistical significance of the resulting autocorrelation. High-significance correlation peaks can be produced by flare-like features in the light curve, well described by red noise processes. To test whether the peaks are a true correlation or due to stochastic red noise processes, we need to quantify the significance of the peaks. For this purpose, we follow the procedure described in detail in \cite{zdcfsign}. Therefore, we simulate time series with the same statistical properties as our observations, i.e. with the same sampling as the light curves used in this work, and the same power spectral density (PSD) and probability density function (PDF), and with no periodicity following the prescription from \cite{simLC}. This method has already been used in previous studies \citep[e.g.][]{zdcfsignej2,zdcfsignej}. Once a large number of random light curves has been simulated (10000 in this work), we calculate the autocorrelation of each curve. Then, assuming a normal distribution of the autocorrelation coefficient in each bin, we estimate the statistical significance, i.e. the probability of obtaining an autocorrelation value due to stochastic red noise random processes. The statistical significance is given as multiples of the standard deviation, considering confidence levels (CL) from 1$\sigma$ to 5$\sigma$, corresponding to probabilities of 68.27, 95.45, 99.73, 99.993666, 99.99994267 per cent. Finally, the uncertainty of the resulting period is estimated making use of equation (3) in \cite{keplerrotation}. 

Other effects should be taken into account as well when using the ZDCF, such as phase and amplitude variations, noise and systematic effects. These factors must be accounted when attempting to determine the true period of the signal \citep[see][]{keplerrotation}. The combination of all these effects can lead to features in the autocorrelation curve like attenuation, alternating high and low peaks due to partial autocorrelation or long-term trends. In this case, the most important are attenuation, phase and amplitude variation, and linear trends and jumps in the light curves. Attenuation and amplitude variation introduce changes of the autocorrelation peak values in the ZDCF (observed for both sources, see the figures in Sections \ref{3cR} and \ref{b2totalmag}), while jumps in the light curve cause long-term trends on the ZDCF  (see Section \ref{3cR}). These effects are present in the analysis; however, they can be distinguished from signatures of periodic variations in the light curves by looking at the ZDCF curve. If these effects are affecting the data series, fluctuations of the autocorrelation coefficient, changes in the amplitude of the peaks and trends in the ZDCF can be observed. A deeper discussion of all these biases is explained in detail in \cite{zdcf}.

\subsection{Lomb-Scargle Periodogram}\label{secLS}
The LS periodogram is one of the most used techniques in periodicity studies of unevenly-sampled time series. This method is based on the discrete Fourier transform, but it differs from the classic periodogram so that the least-square fit of the sine functions is minimized \citep{lomb1976, scargle1982, vanderplas}. This modification of the classic periodogram accounts for the problem of the irregularly-sampled time series, avoiding interpolation and the appearance of spurious peaks in the spectrum \citep{radio1156}.

In order to compute and quantify the significance of the resulting peaks of the periodograms, we follow the procedure described by \cite{LSsignificance}. Red noise processes can lead to spurious peaks and large fluctuations in the periodogram, specially at low frequencies (long timescales). Assuming that the periodogram follows a power-law function model of the form $P(f)=Nf^{-\alpha}$, then the logarithm of the periodogram can be expressed as a linear function. Under these considerations, we can estimate the red noise continuum by fitting the logarithm of the periodogram with a linear function. With this estimator we can formulate the null hypothesis and calculate the probability that the data were produced by chance by a non-periodic signal. Adopting this method, the confidence levels in the periodogram also adopt a power law shape. This power law fit was also compared with the PSD estimation of the data to check their compatibility. All the fitted functions were found to be compatible with the derived PSDs of the light curves.

The uncertainty in the period determined with the LS is commonly given by the full width half maximum (FWHM) of a Gaussian fit centered at the position of the maximum power of the corresponding peak \citep[see][]{wwzex, zhang2017a, zhang2017b}. This value only gives a rough estimation of the period uncertainty. For a more robust estimation, effects like spurious background peaks or aliases should be taken into account \citep{vanderplas}.

When using the LS periodogram, it is important to consider the caveats and limitations of the method. First, it is important to choose a proper frequency grid so as not to miss relevant information and to avoid a large computational burden. Also, one must take into account the possible appearance of spurious peaks due to harmonics of the true frequency and other kind of aliases that can introduce a sequence of false peaks, caused for example by annual observation patterns that can introduce a spike at a period of 1 year \citep[see][]{vanderplas}. All the effects commented in Section \ref{secZDCF} also have their impact on the LS periodogram, leading to effects such as splitting or displacement of the true peaks or big bumps at a false frequency. Regarding the most important effects for our data, attenuation and amplitude variations cause the periodogram to produce two peaks on both sides on the correct period. Linear trends and jumps in the light curve introduce spurious peaks in the periodogram at large periods, which can lead to the wrong identification of the true period \citep{keplerrotation}. In our analysis, we checked that the detected peaks do not correspond to harmonic peaks or sampling effects.

\subsection{Weighted Wavelet Z-Transform}\label{secWWZ}
Alternatively to the traditional Fourier techniques for periodicity analysis, the wavelet transform is another widely used method due to its localization ability in both time and frequency domains \citep{wavelet_theory}, decomposing a time series into time-frequency space, finding the dominant modes of variability and their evolution with time. It is a powerful tool used for AGN variability studies \citep[see][]{radio1156, wwz_ex, wwzex}.

An alternative and more suitable form of the wavelet technique for unevenly sampled data, the WWZ, is used in this work. It was developed to improve its performance under these conditions \citep{wwz}, based on the Morlet wavelet \citep{morlet_wavelet, wavelet2}. This variation from the original method makes use of the z-transforma and improves the performance of the wavelet transform under realistic conditions of unevenly sampled data series \citep{wavelet}. Additionally, the PSD of the WWZ is calculated, giving the same information as the WWZ for a periodogram.

For an estimate of the significance we follow a procedure similar to the one used for the LS periodogram (Section \ref{secLS}). We simulate 10000 non-periodic light curves with the same sampling as our data. For each of these light curves, we compute the WWZ and the corresponding PSD. The significance is then calculated using the mean PSD of the simulated light curves as the non-periodic baseline following the method described in \cite{LSsignificance}. Again, in the same way as in LS periodogram, the period uncertainty is given by the FWHM of a Gaussian fit of the corresponding peak in the PSD. The same considerations we explained for the uncertainty estimation in the LS must be taken into account for the WWZ in order to make a more precise estimation of this value.

The main caveats and considerations when using the WWZ method are the choice of the wavelet function (width and shape of the function), the adequate time and frequency resolution, the choice of scales in order to have an optimal sampling, and the region in which edge effects become relevant due to the finite length of the time series, known as the cone of influence (COI), which can lead to the appearance of spurious peaks in the power spectrum of the WWZ \citep[see][]{wavelet_theory}. An optimal sampling and scale were selected to perform this analysis, searching for a balance between the frequency resolution and computational time needed. Also, the effect of the COI was taken into account when calculating the WWZ, PSD and significance of the data.

\section{Results}\label{results}
The periodicity analysis has been applied to the optical total flux in R and V bands, its polarization and the $\gamma$-ray light curves, using the methods described in Section \ref{per_methods}. The results for 3C 66A and B2 1633+38 are presented in Section \ref{res3c66a} and Section \ref{resb2}, respectively. The final results for each source are reported in Table \ref{results_all}.

\subsection{3C 66A}\label{res3c66a}
\subsubsection{Total optical magnitude}\label{3cR}
\begin{figure*}
\centering
\subfigure{\includegraphics[width=0.85\columnwidth]{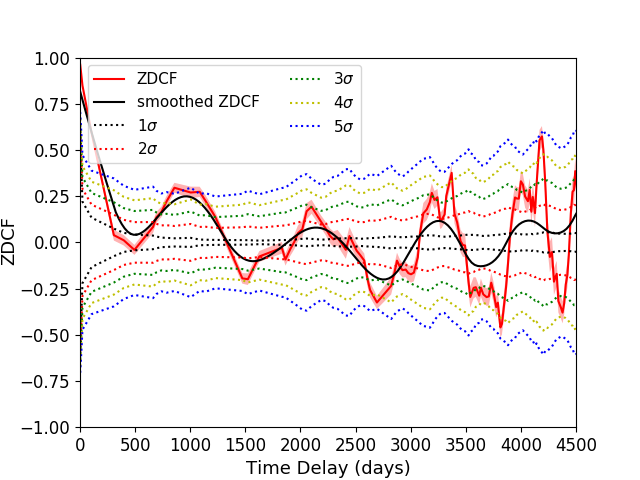}}
\subfigure{\includegraphics[width=0.85\columnwidth]{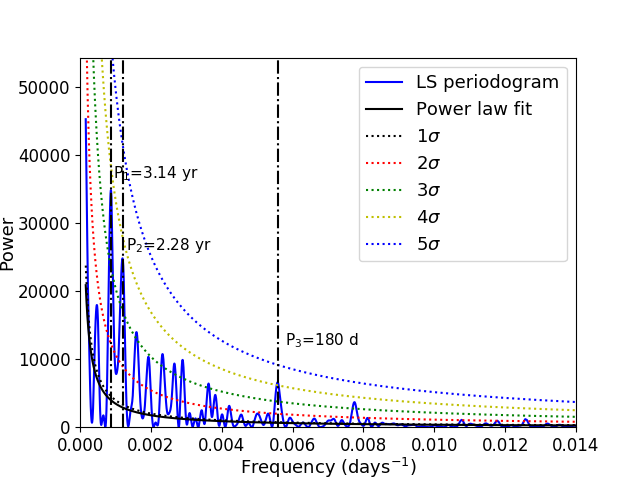}}
\subfigure{\includegraphics[width=0.85\textwidth]{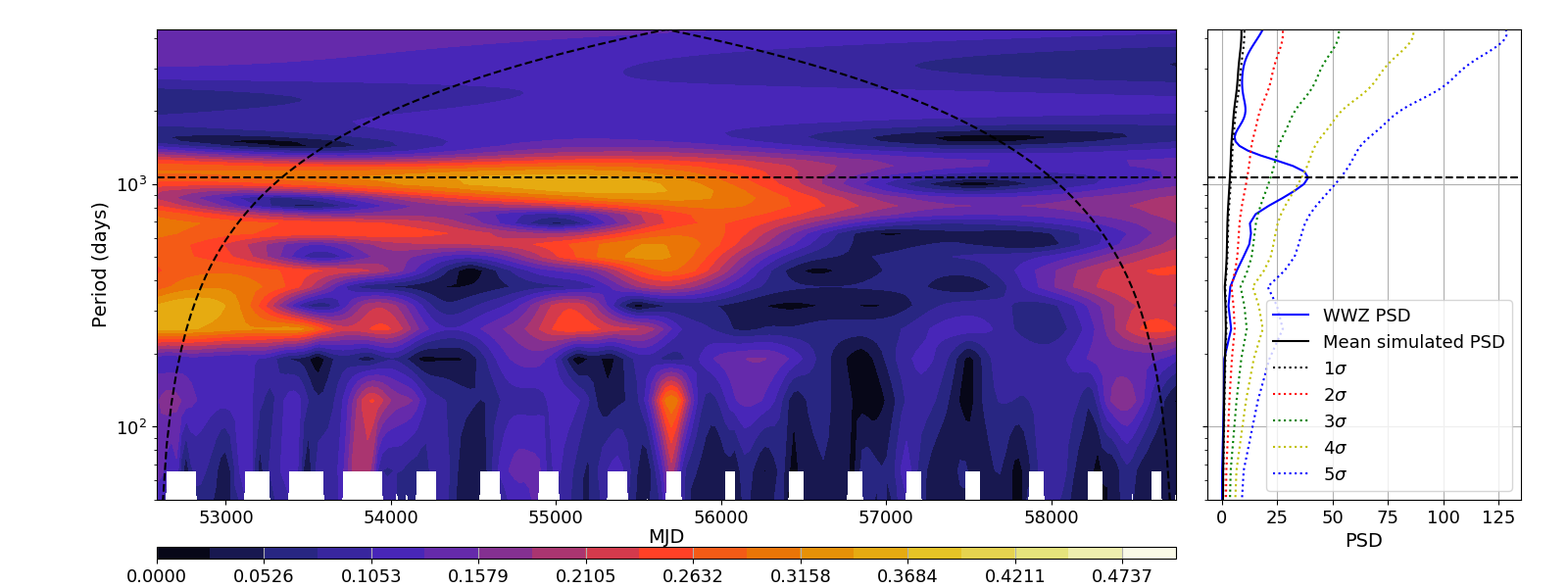}}
\caption{Periodicity analysis of the R band data of 3C 66A. \textit{Top left}: ZDCF method. The autocorrelation curve is given in red and the smoothed curve in black. \textit{Top right}: Lomb-Scargle periodogram. The periodogram is given in blue and the power law fit in black. \textit{Bottom}: WWZ method. The left panel shows the 2D power spectrum as a function of period and time. The black dashed curve represents the COI. The right panel shows the PSD in blue and the mean simulated PSD in black. The coloured dotted lines represent the different significance levels. The black horizontal dashed line marks the peak of the PSD.}
\label{3c66a_all_plots}
\end{figure*}

We analyse the R and V band light curves of 3C 66A with the three different methods presented in Section \ref{per_methods}. We perform the analysis in magnitude scale instead of flux scale, since the latter is more biased towards large flares. Because of the better sampling we focus on the results of the R band light curve. The V band light curve shows a largely uneven sampling and a poor time coverage in comparison with the R band ($\sim$50 observations per year and less than 700 total observations for the V band light curve, compared with more than 250 observations per year and more than 4000 observations in the case of the R band light curve of this target). The results of the V band light curve are presented in Fig. \ref{3c66a_all_plots_V}. The results of the R band analysis are shown in Fig. \ref{3c66a_all_plots}. 

The ZDCF in the R band for 3C 66A shows a clear sinusoidal shape as expected in case of a periodic behaviour, with a period of 2.98$\pm$0.14 years. The associated uncertainty is calculated using equation (3) in \cite{keplerrotation}. The significance of the peaks in the ZDCF varies between the 3$\sigma$ and 5$\sigma$ CL. Flare-like and short-term variability features are reflected in the curve as small fluctuations in the autocorrelation coefficient.

The resulting LS periodogram exhibits a prominent peak centered at a period value of 3.14$\pm$0.24 years, at a significance of $\sim$4$\sigma$. While the significance is lower than that obtained with the ZDCF method, the measured period is consistent. The periodogram also shows a secondary peak centered at a period of 2.28 years. The existence of two peaks can be interpreted as a variation of the timescale associated to two different activity states of the source. This can be tested by dividing the R band light curve in two parts: the first one before MJD 55700, which corresponds to a possible high state of the source, with higher flux, in which a period of $\sim$3 years is detected; and a second part from MJD 55700 onwards, which shows a possible low state of the source, and a period of $\sim$2.3 years. The peak at a period of 2.28 years is compatible with the results of the V band light curve periodicity analysis shown in Fig. \ref{3c66a_all_plots_V}. This latter value is also consistent with the possible long-term quasi-periodicities found by \cite{belokon2003} and \cite{kaur2017}. Furthermore, a peak above the 4$\sigma$ CL is found at a period of $\sim$180 days. This peak is compatible with the period detected by \cite{fan2018} with a value of 156$\pm$17 days in the R band, and 156$\pm$15 days in the I band.

The color-scaled histogram of the WWZ shows a periodic feature, also reflected on the PSD, centered at 2.92$\pm$0.79 years at $\sim$4$\sigma$ CL. This result is consistent with those derived from the ZDCF and the LS methods. Non-significant peaks due to flare-like short-term variability can be identified in the PSD curve, and can also be seen in the color-scaled power representation of the WWZ at lower values of the period. At long periods, the effect of the red noise becomes more important, introducing a bump above $\sim$6 years. The V-band WWZ presents only the secondary peak, centered at 2.27$\pm$0.47 years at a CL of $\sim$3$\sigma$. Since almost all of the V-band data were obtained during the relatively fainter epoch, all the previous results are consistent.

\subsubsection{Polarized optical magnitude}\label{3cpolR}
\begin{figure*}
\centering
\subfigure{\includegraphics[width=0.85\columnwidth]{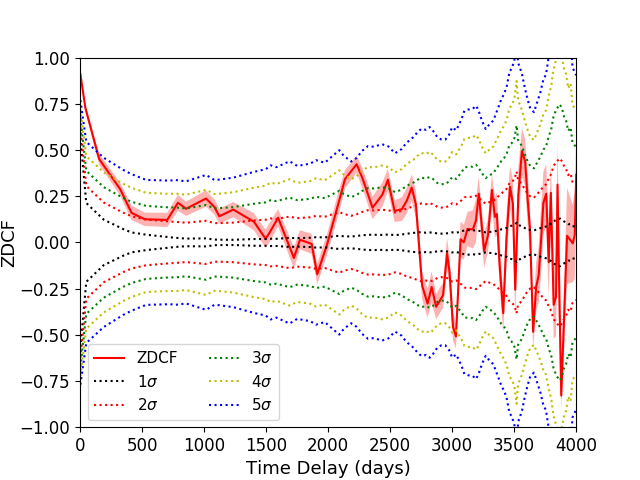}}
\subfigure{\includegraphics[width=0.85\columnwidth]{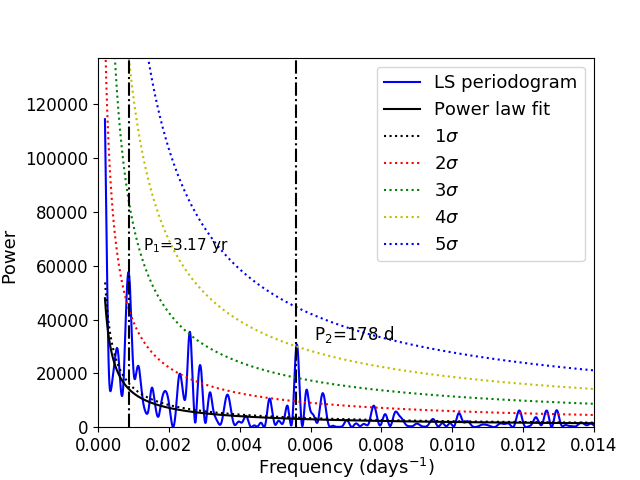}}
\subfigure{\includegraphics[width=0.85\textwidth]{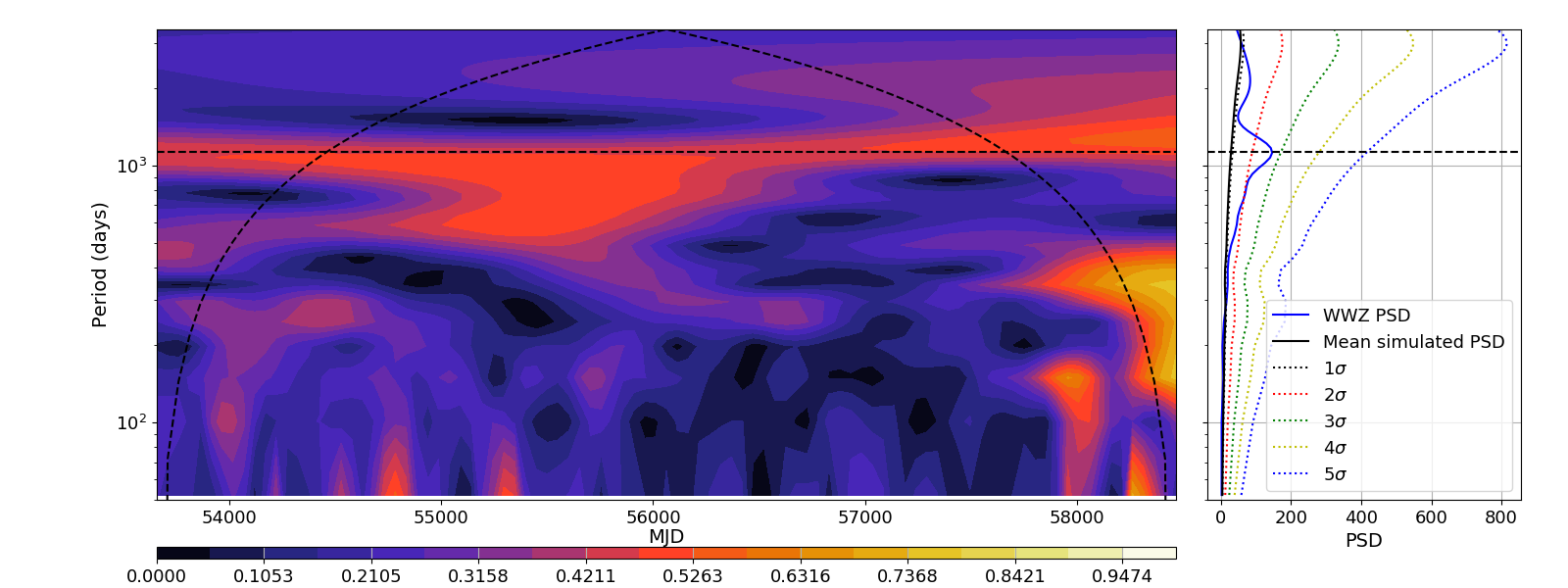}}
\caption{Periodicity analysis of the polarized R band data of 3C 66A. Same description as Fig. \ref{3c66a_all_plots}.} 
\label{3c66a_all_plots_pol}
\end{figure*}

The periodicity analysis was performed on the polarized data provided by the Steward observatory, the polarimetric observations of the St. Petesburg University made in St. Petersburg and the Crimean Astrophysical Observatory. The results are presented in Fig. \ref{3c66a_all_plots_pol}. 3C 66A shows no hints for periodic behaviour with any of the used techniques in this work. The ZDCF has no periodic structure, with no significant peaks that can be associated to a periodic emission. A strong fluctuation at a time delay of $\sim$2000 days appears, but since there is no periodic structure in the curve and no other method agrees with this peak, it is likely due to a non-periodic emission process. Also, at such high time delays, only a small part of the light curve is being compared, so it is more probable to find randomly high autocorrelation values. 
The LS periodogram and the wavelet analysis show no significant long-term variability peaks reaching a significance of 3$\sigma$ around the period seen for the total magnitude. However, a shorter-period peak appears in the LS with a period of 178 days at a CL of $\sim$4$\sigma$. This peak is consistent with the one seen in the total flux periodogram with a period of 180 days.

\begin{figure}
\includegraphics[width=\columnwidth]{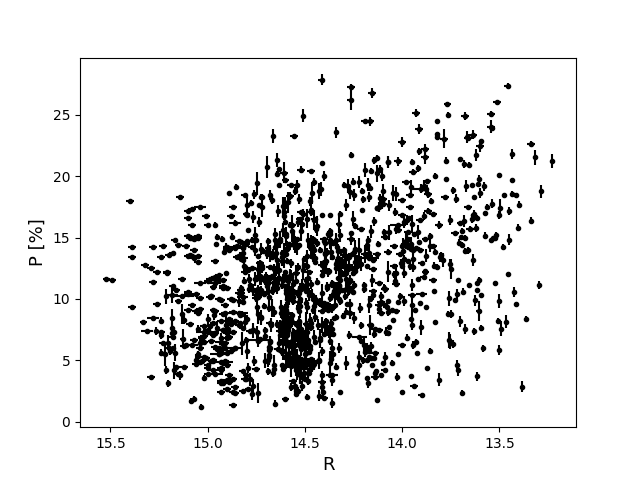}
\caption{The observed polarization degree against the R band total magnitude for 3C 66A. These two quantities have a Spearman's correlation coefficient of $\rho=0.17$.}
\label{corr_pol_R_3c66a}
\end{figure}

This absence of long-term periodicity in the optical polarized light curve of 3C 66A is in agreement with the lack of correlation observed between the optical magnitude and the polarization degree. This result is shown in Fig. \ref{corr_pol_R_3c66a}. There is no correlated behaviour between the magnitude and the polarization of this source. We performed an estimation of the Spearman's rank correlation coefficient ($\rho$) to quantify the degree of (non-)~correlation observed in 3C 66A, obtaining a value of $\rho=0.17$, $p=0.0017$ between the polarization degree and the R band magnitude, where $p$ is the probability for null hypothesis.

\subsubsection{$\gamma$-ray flux}
\begin{figure*}
\centering
\subfigure{\includegraphics[width=0.85\columnwidth]{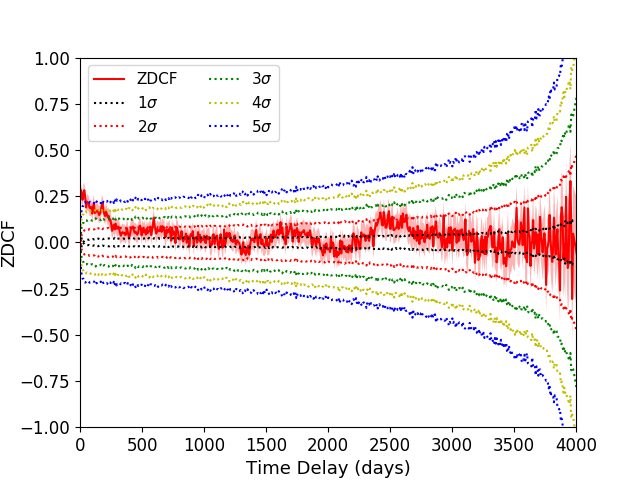}}
\subfigure{\includegraphics[width=0.85\columnwidth]{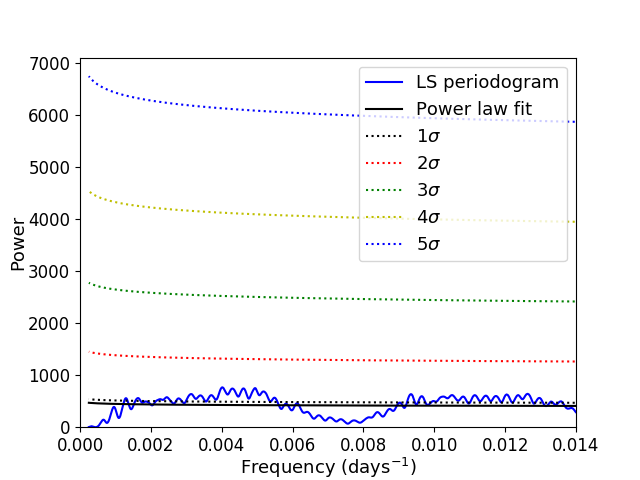}}
\subfigure{\includegraphics[width=0.85\textwidth]{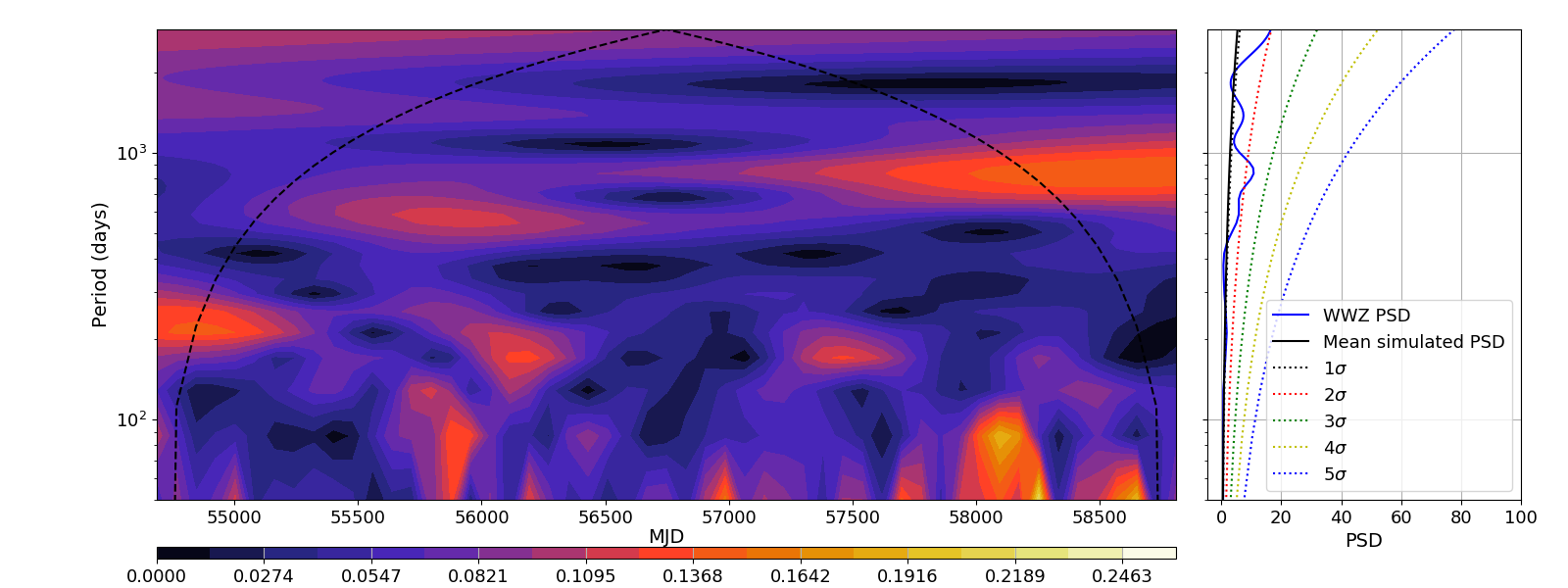}}
\caption{Periodicity analysis of the $\gamma$-ray flux for 3C 66A. Same description as Fig. \ref{3c66a_all_plots}.} 
\label{3c66a_all_plots_gamma}
\end{figure*}

We have also searched for possible periodicity of this source in the high energy $\gamma$-ray regime with the light curve extracted from the \textit{Fermi}-LAT data. All three methods agree that there is no indication of long-term periodicity in the $\gamma$-ray emission of 3C 66A. There is no periodic structure or fluctuation in the ZDCF and no significant peaks in the periodogram or the wavelet transform (Fig. \ref{3c66a_all_plots_gamma}).

\subsection{B2 1633+38}\label{resb2}
\subsubsection{Total optical magnitude}\label{b2totalmag}
\begin{figure*}
\centering
\vspace{-5mm}
\subfigure{\includegraphics[width=0.84\columnwidth]{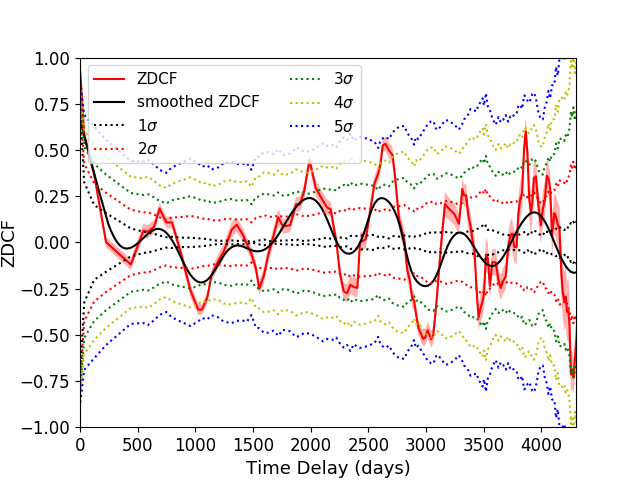}}
\vspace{-4mm}
\subfigure{\includegraphics[width=0.84\columnwidth]{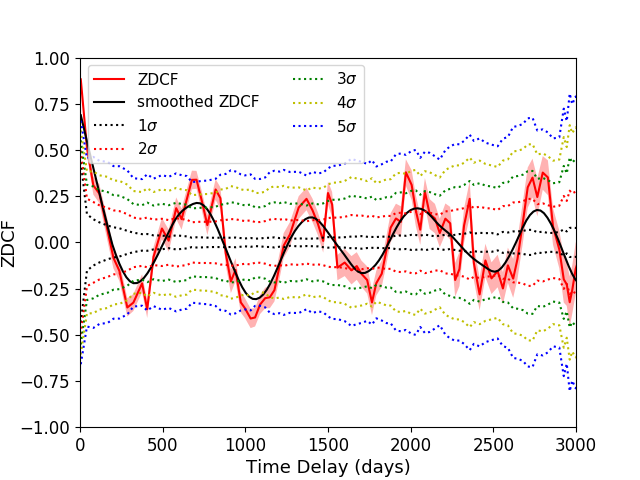}}
\subfigure{\includegraphics[width=0.84\columnwidth]{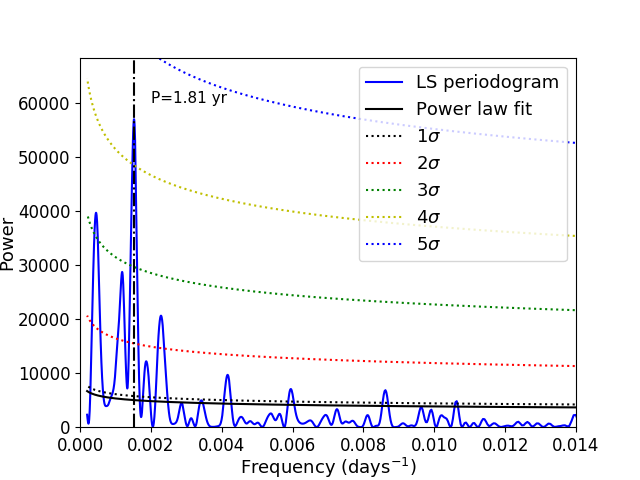}}
\vspace{-4mm}
\subfigure{\includegraphics[width=0.84\columnwidth]{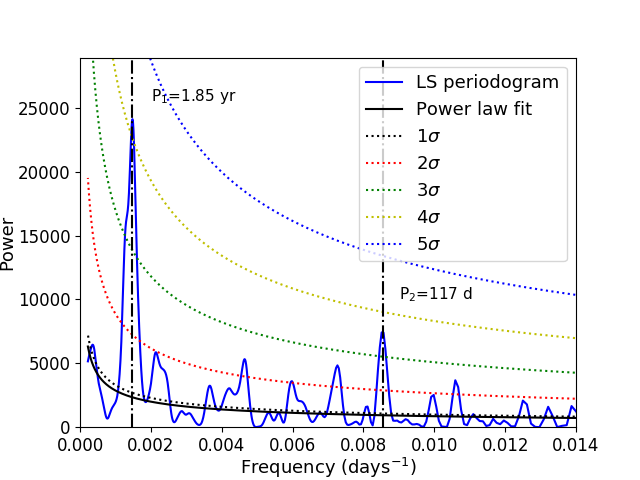}}
\subfigure{\includegraphics[width=0.83\textwidth]{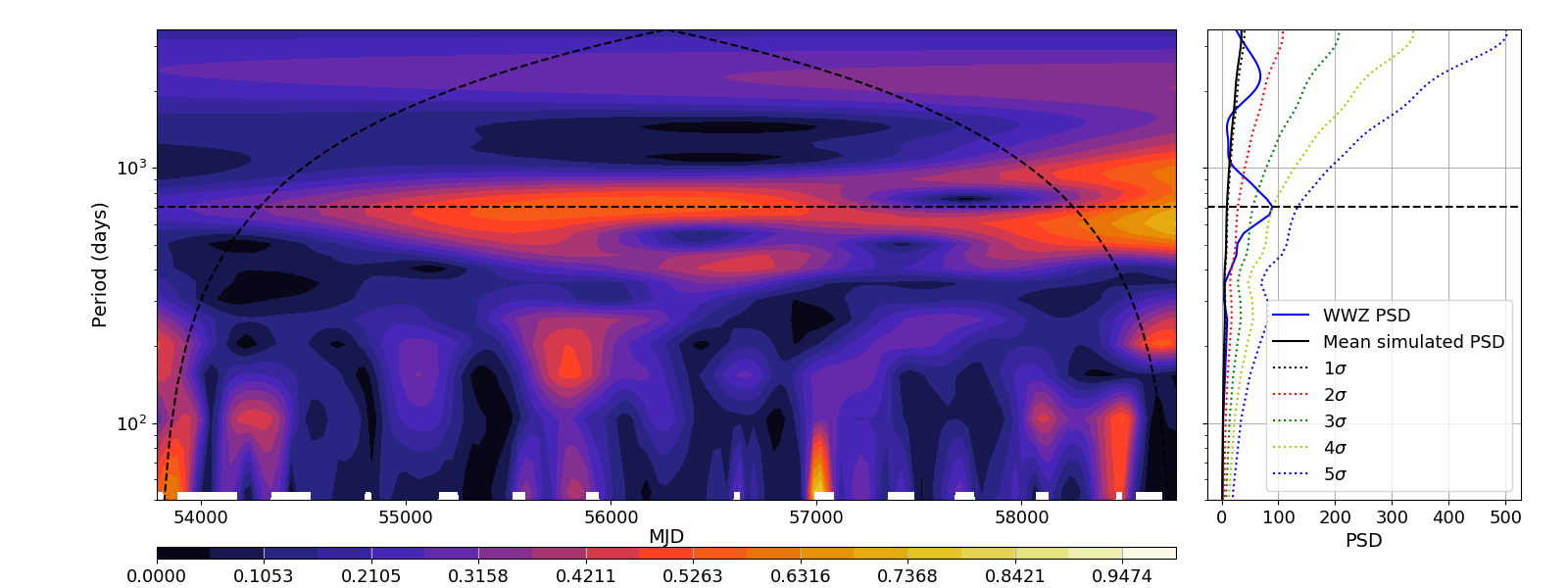}}
\subfigure{\includegraphics[width=0.83\textwidth]{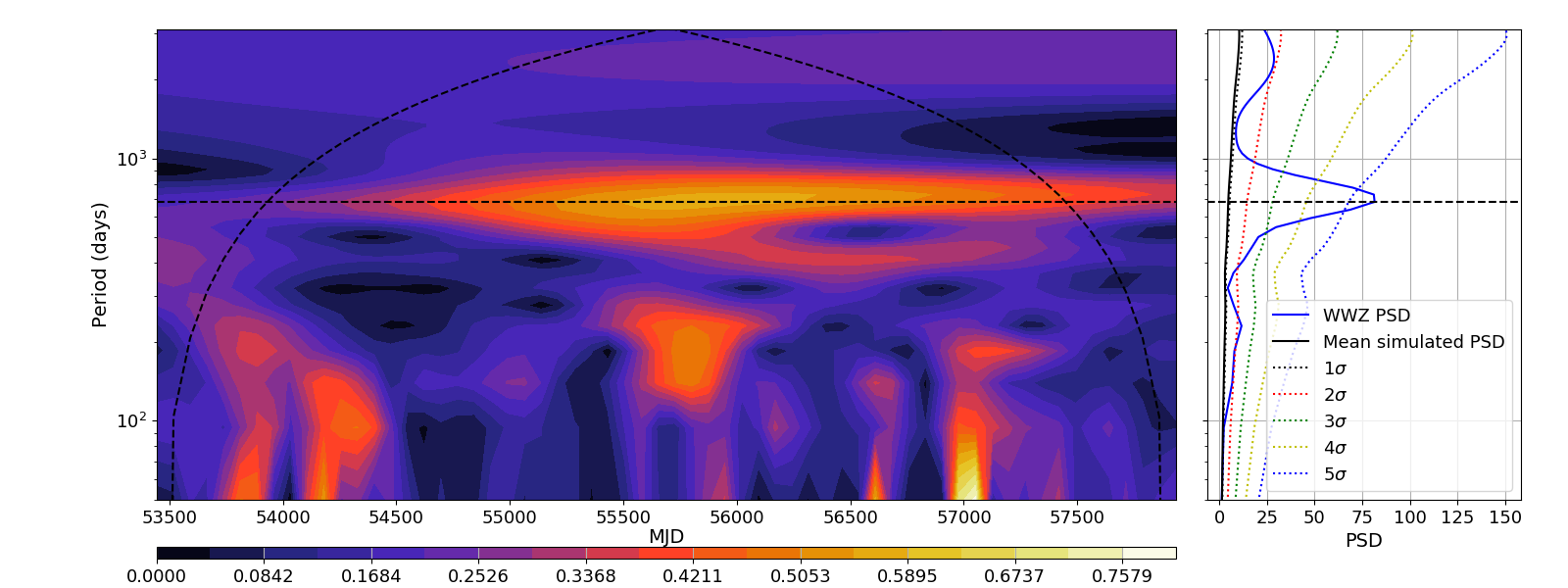}}
\caption{Periodicity analysis of B2 1633+38. \textit{First row}: ZDCF method in R (left) and V (right) bands. The autocorrelation curve is given in red and the smoothed curve in black. \textit{Second row}: Lomb-Scargle periodogram in R(left) and V (right) bands. The periodogram is given in blue and the power law fit in black. \textit{Third row}: WWZ diagram in the R band. \textit{Fourth row}: WWZ diagram in the V band. The left panels show the 2D power spectrum as a function of period and time. The black dashed curves represent the COI. The right panels show the PSD in blue and the mean simulated PSD in black. The coloured dotted lines represent the different significance levels. The black horizontal dashed lines mark the peak of the PSD.}
\label{b2_all_plots}
\end{figure*}

In the case of B2 1633+38, we are presenting both R and V bands, since both have a good temporal coverage and sampling, as shown in Fig. \ref{b2LC}. The results of all three methods are shown in Fig. \ref{b2_all_plots}. The ZDCF presents a sinusoidal shape characteristic of a periodic component to its optical emission. The measured period with this method has a value of 1.91$\pm$0.05 years for the V band and 1.80$\pm$0.04 years for the R band. The maxima and minima of the curve reach significances between 2$\sigma$ and 5$\sigma$. Flare-like features on the light curve can be seen in the ZDCF as fluctuations in the autocorrelation coefficient, as is seen for 3C 66A.

The LS periodograms peak at a period of 1.85$\pm$0.17 years in the V band and at 1.80$\pm$0.10 years in the R band. These values are compatible with those calculated with the autocorrelation method within their uncertainties. The significance of the peaks reaches values slightly higher than 4$\sigma$ in both cases. 
Also, a secondary peak (above 3$\sigma$ CL) with a period of 117~days is found in the V-band periodogram.

Regarding the WWZ for this source there is also an agreement with the two other methods. The measured values of the period with the wavelet transform are 1.93$\pm$0.40 years in the V band, with a significance above 5$\sigma$, and 1.89$\pm$0.48 years in the R band, at the 4$\sigma$ CL. The results of all three methods for the analysis of B2 1633+38 can be seen in Table \ref{results_all} for both filters, with their corresponding uncertainties and significance levels.

\subsubsection{Polarized optical magnitude}
\begin{figure*}
\centering
\subfigure{\includegraphics[width=0.85\columnwidth]{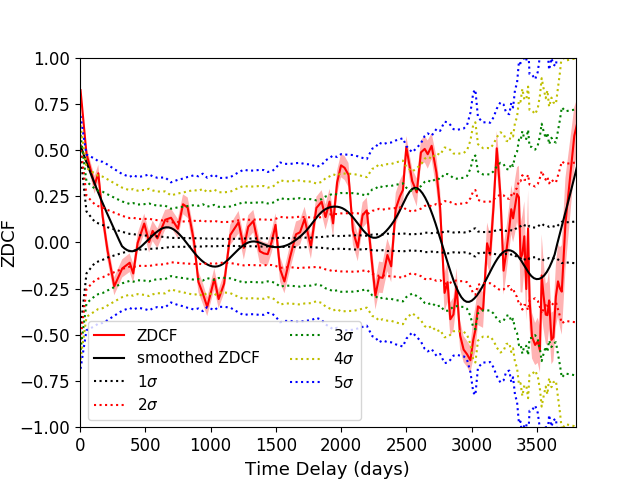}}
\subfigure{\includegraphics[width=0.85\columnwidth]{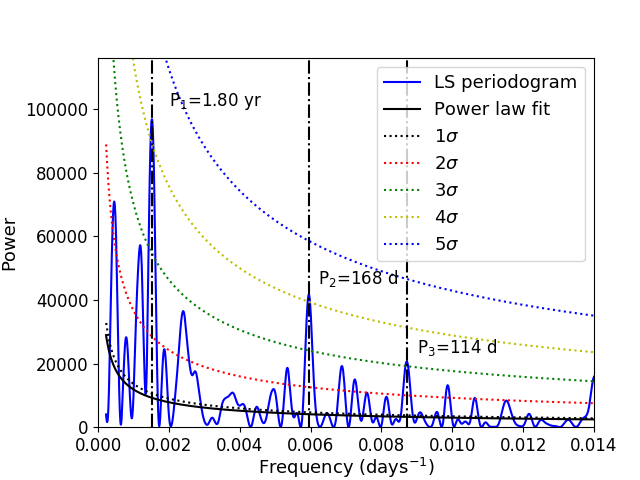}}
\subfigure{\includegraphics[width=0.85\textwidth]{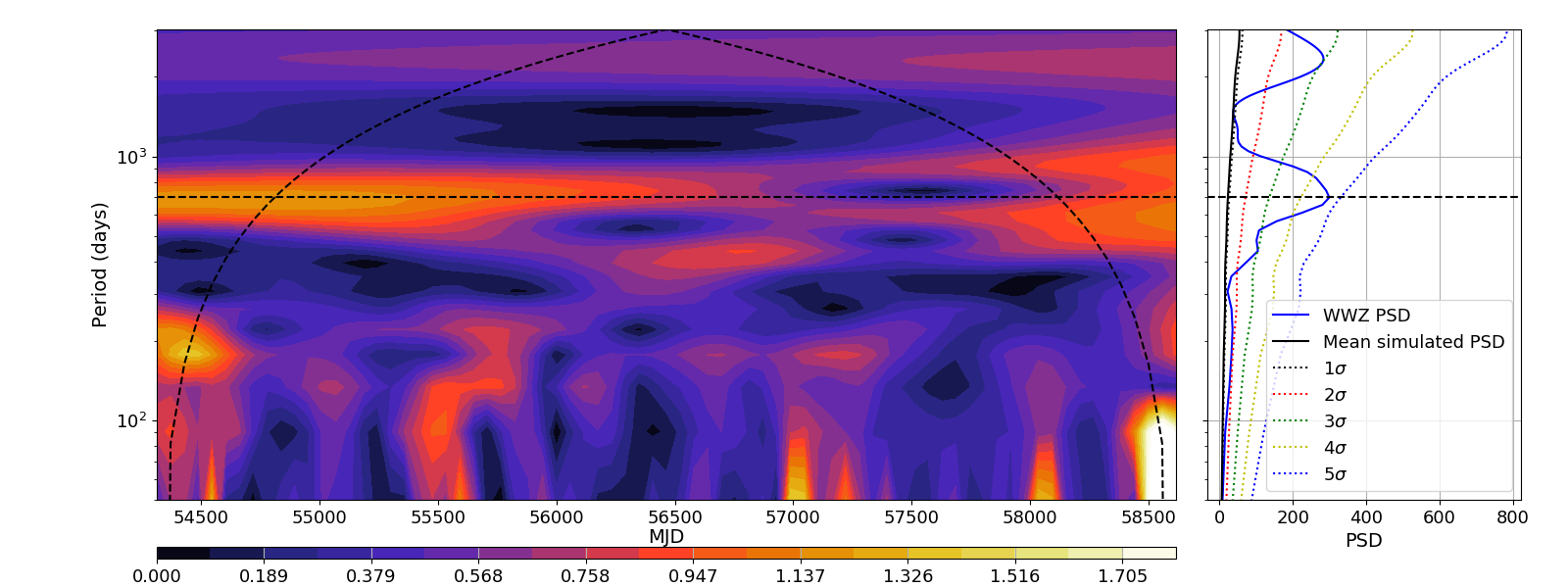}}
\caption{Periodicity analysis of the polarized R band data of B2 1633+38. Same description as Fig. \ref{3c66a_all_plots}.} 
\label{b2_all_plots_pol}
\end{figure*}

\begin{figure}
\includegraphics[width=\columnwidth]{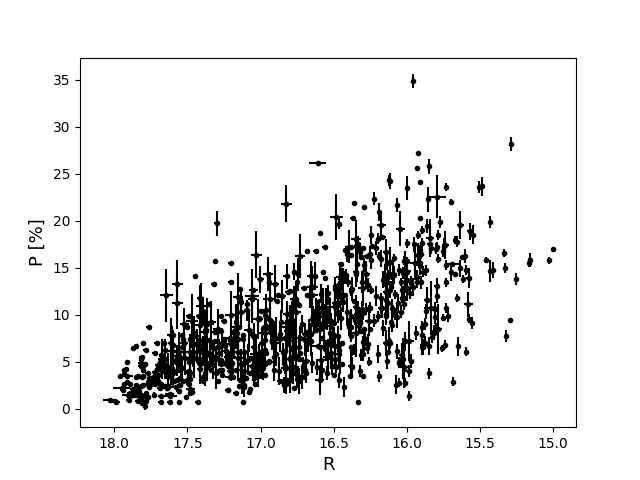}
\caption{The observed polarization degree against the R band total magnitude for B2 1633+38. These two quantities have a Spearman's correlation coefficient of $\rho=0.66$.}
\label{corr_pol_R_b2_1633}
\end{figure}

Contrary to what has been seen on 3C 66A, this source does exhibit a possible periodic variability also in the polarized magnitude light curve. The results of the analysis are shown in Fig. \ref{b2_all_plots_pol}. The ZDCF curve shows the same behaviour as for the total flux, having a sinusoidal shape with alternating maxima and minima at CL between $\sim$2$\sigma$ and almost 5$\sigma$. The variability timescale measured with this method is 1.79$\pm$0.06 years. The LS periodogram reveals a prominent peak at a period value of 1.81$\pm$0.12 years with a significance of $\sim$4$\sigma$, compatible with both the ZDCF and the total flux case. Finally, the WWZ and its corresponding PSD estimate a period of 1.92$\pm$0.57 years above the 4$\sigma$ CL in agreement with the two other methods. This result is expected on the basis of the strong correlation between continuum flux and polarization degree found by \cite{raiteri2012}. This correlated behaviour between the polarization and the flux can be seen in Fig. \ref{corr_pol_R_b2_1633}. The polarization degree shows a positive correlation with the total optical magnitude for B2 1633+38 with a Spearman's correlation coefficient of $\rho=0.66$, $p<0.001$. 
In addition, hints of periods on the order of $\sim$100 days at CL of $\sim$3$\sigma$ are found in the LS periodogram for both total and polarized optical flux, and also a peak with a period of $\sim$170 days at CL of $\sim$4$\sigma$ in the polarized magnitude periodogram.

\subsubsection{$\gamma$-ray flux}
\begin{figure*}
\centering
\subfigure{\includegraphics[width=0.85\columnwidth]{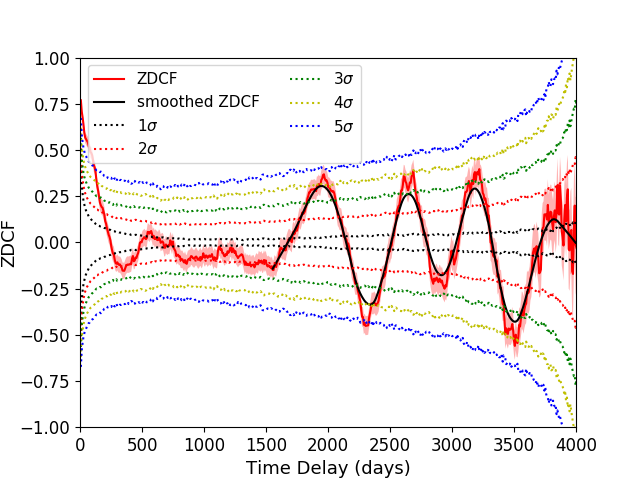}}
\subfigure{\includegraphics[width=0.85\columnwidth]{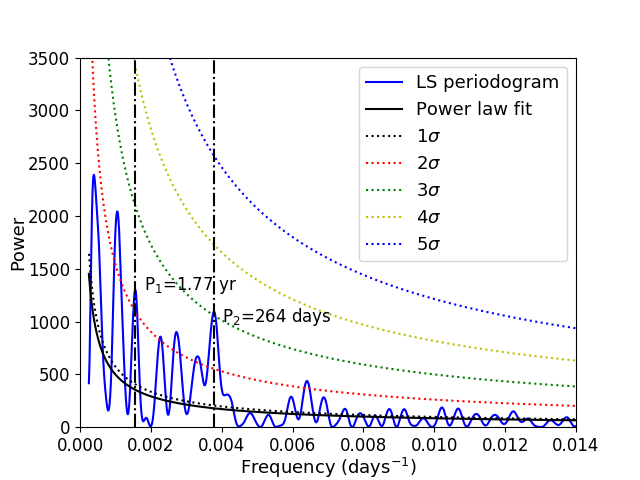}}
\subfigure{\includegraphics[width=0.85\textwidth]{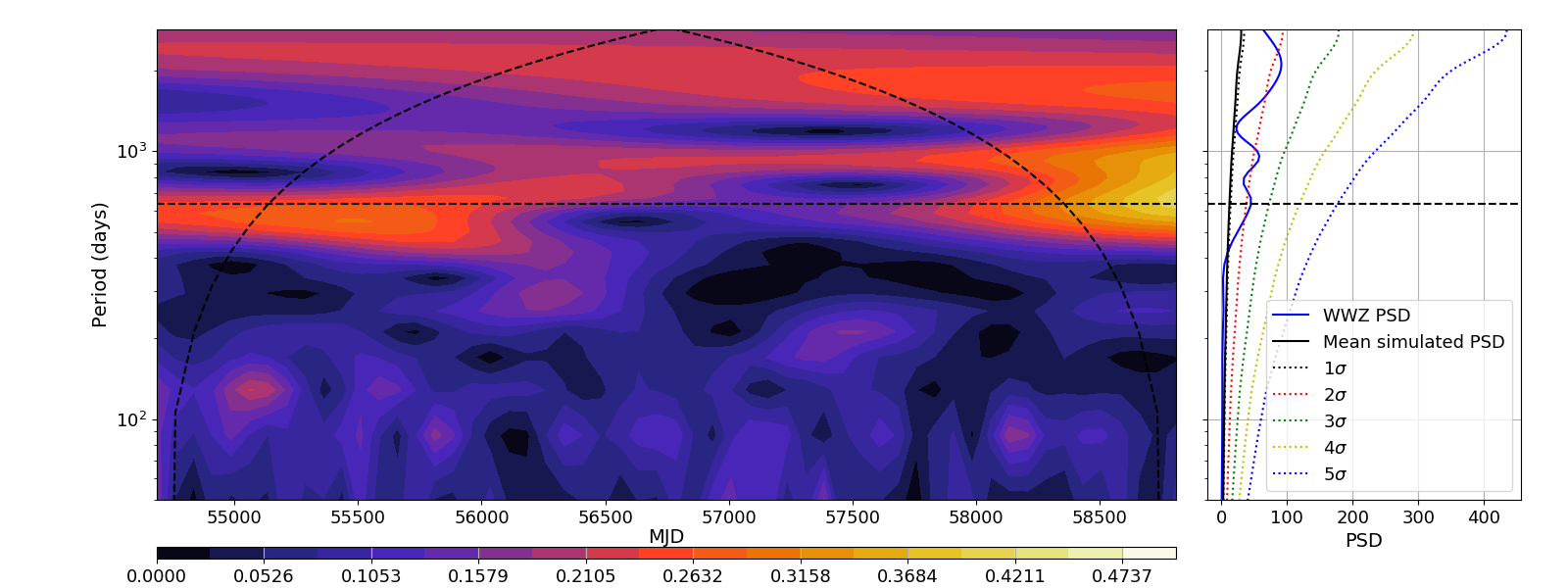}}
\caption{Periodicity analysis of the $\gamma$-ray flux for B2 1633+38. \textit{Top left}: ZDCF method. The autocorrelation curve is given in red and the smoothed curve in black. \textit{Top right}: Lomb-Scargle periodogram. The periodogram is given in blue and the power law fit in black. \textit{Bottom}: WWZ diagram. The left panel shows the 2D power spectrum as a function of period and time. The black dashed curve represents the COI. The right panel shows the PSD in blue and the mean simulated PSD in black. The coloured dotted lines represent the different significance levels. The black horizontal dashed line marks the peak of the PSD.}
\label{b2_all_plots_gamma}
\end{figure*}

The $\gamma$-ray variability analysis, as it happened with the optical total and polarized magnitude light curve analysis, shows a hint of periodic variability on the timescale of $\sim$2 years. The results of the $\gamma$-ray band analysis are presented in Fig. \ref{b2_all_plots_gamma}. The ZDCF shows a series of maxima and minima with a periodicity of 1.59$\pm$0.07 years and significances of $\sim$3-4$\sigma$ with a periodic structure starting at a time lag of 1500 days. Up to that delay, the ZDCF shows no prominent peak, with an autocorrelation coefficient close to zero. This can be explained by observing the corresponding light curve. During the first 2000 days there were three major outbursts, and afterwards the $\gamma$-ray flux seems to be in a quiescent state that finishes at MJD 58000, when another outburst commences. This quiescent state may be the reason why an autocorrelation value close to zero at delays shorter than 1500 days is observed. It suggests that there is no real periodic behaviour, but a series of unrelated and equidistant flares. However, we cannot exclude the possibility that the limited sensitivity of \textit{Fermi}-LAT prevented us from detecting lower-amplitude periodic flares during the quiescent period. In this case, we are only smoothing the part of the ZDCF with the prominent peaks to estimate time lags and derive the period value. A peak in both the LS periodogram and the PSD of the wavelet transform with approximately the same CL and consistent period values appears to support a possible periodic variation (1.77$\pm$0.12 years with the periodogram and 1.75$\pm$0.48 years with the wavelet transform). These results are not significant enough to claim for a solid periodic detection, however \cite{raiteri2012} found a strong correlation between optical and $\gamma$-ray flares. Despite that, flux ratios change with time and there is a general trend for the optical variations to follow those observed in the $\gamma$-rays by few days. This fact gives marginal support to our finding of quasi-periodic variations at high energies. Two more peaks appear at higher periods ($\sim$3 years and $\sim$6 years). The 3 year peak seems incompatible with the results of the ZDCF and the optical curves, and also inconsistent with time lags between flares in the \textit{Fermi}-LAT light curve, so it is not considered as possible period. The 6 year peak is identified as red noise. A peak with a period value of $\sim$264 days is also identified in the LS periodogram at a CL of 3$\sigma$ that could be indicative of a type of variability in shorter timescales.

\begin{table*}
\centering
\caption{Results of the periodicity analysis of 3C 66A and B2 1633+38 for the R and V bands in both the total and polarized flux, and the high energy $\gamma$-ray flux.}
\label{results_all}
\begin{tabular}{cccccc}
\hline
Source   & Band & Method & Period (Years) & Error (Years) & Significance \\ \hline
\multirow{13}{*}{3C 66A} & \multirow{4}{*}{R band}          & ZDCF &     2.98   &   0.14    &  3$\sigma$-5$\sigma$ \\ \cline{3-6} 
  & & \multirow{2}{*}{LS}  &    3.14    &   0.24    & $\sim$4$\sigma$ \\ \cline{4-6} 
  & &      &   2.28     &   0.15    & >3$\sigma$  \\ \cline{3-6} 
  & & WWZ &    2.92    &    0.79   & $\sim$4$\sigma$  \\ \cline{2-6} 
  & \multirow{3}{*}{V band}          & ZDCF &    2.04    &    0.10   & 2$\sigma$-3$\sigma$ \\ \cline{3-6} 
  & & LS   &    2.48    &   0.29    & $\sim$2$\sigma$ \\ \cline{3-6} 
  & & WWZ  &   2.28     &   0.47    & $\sim$3$\sigma$ \\ \cline{2-6} 
  & \multirow{3}{*}{Polarized R band}      & ZDCF &    --    &   --    & -- \\ \cline{3-6} 
  & & LS   &    3.17    &    0.37   & <3$\sigma$ \\ \cline{3-6} 
  &   & WWZ  &   3.07     &    0.85   &  <3$\sigma$ \\ \cline{2-6} 
  & \multirow{3}{*}{Gamma rays} & ZDCF  &  --  & -- & -- \\ \cline{3-6} 
  &   & LS   &   --     &   --    &  -- \\ \cline{3-6} 
  &   & WWZ   &   --    & -- &  -- \\ \hline
  \multirow{12}{*}{B2 1633+38} & \multirow{3}{*}{R band}          & ZDCF   &   1.80     &  0.04 &  2$\sigma$-5$\sigma$ \\ \cline{3-6} 
  &   & LS     &    1.80    &   0.10    &      >4$\sigma$        \\ \cline{3-6} 
  &   & WWZ    &    1.89    &   0.48    &       $\sim$4$\sigma$       \\ \cline{2-6} 
  & \multirow{3}{*}{V band}   & ZDCF   &    1.91    &       0.05&   2$\sigma$-5$\sigma$           \\ \cline{3-6} 
  &   & LS     &   1.85     &   0.17    &       >4$\sigma$       \\ \cline{3-6}
  &   & WWZ    &   1.93     &   0.40    &      >5$\sigma$        \\ \cline{2-6} 
  & \multirow{3}{*}{Polarized R band}      & ZDCF   &   1.79     &  0.06     &       2$\sigma$->4$\sigma$       \\ \cline{3-6} 
  &   & LS     &    1.81    &    0.12   &      >4$\sigma$        \\ \cline{3-6} 
  &   & WWZ    &    1.92    &   0.57    &    >4$\sigma$  \\ \cline{2-6} 
  & \multirow{3}{*}{Gamma rays} & ZDCF   &    1.59    & 0.07      &   3$\sigma$-4$\sigma$           \\ \cline{3-6} 
  &   & LS     &   1.77     &   0.12    &        $\sim$2$\sigma$      \\ \cline{3-6} 
  &   & WWZ    &   1.75     &   0.48    &        $\sim$2$\sigma$      \\ \hline
\end{tabular}
\end{table*}

\section{Discussion}\label{discussion}
After the analysis of both the total and polarized optical magnitude light curves, we can highlight the different behaviour observed in these two sources. While 3C 66A shows a clear period only in the total optical light curve, B2 1633+38 behaves periodically for both total and polarized light. This difference can also be appreciated in the high energy $\gamma$-ray regime. 3C 66A shows no long-term variability, with variations only at short timescales. Moreover, the high energy emission of B2 1633+38 exhibits a hint of a periodic flux variability. However, the confirmation of QPO in $\gamma$-ray with \textit{Fermi}-LAT data, previously claimed by other authors, has been questioned by \cite{Covino2019}.

Year-like timescale quasi-periodicities shown in blazars have been interpreted within various scenarios \citep[e.g.][]{Ackermann15, sandrinelli2016}, and can be grouped into two general cases:
\begin{enumerate}
    \item The existence of a supermassive binary black hole (SBBH), as proposed for OJ 287 \citep{Sillanpaa88, lehto1996, Villata99};
    or PG 1302-102 \citep{Graham15}. The presence of a second perturbing black hole could either modulate 
    the accretion rate or produce a precession in the jet(s). Models based on a SBBH system have been 
    proposed \citep{Cavaliere17, Caproni17, Sobacchi17} and applied to the case of PG 1553+113 \citep{Ackermann15}. \cite{Sobacchi17} proposed a simple spine-sheath precessing jet model, which is able to reproduce the recurrence observed in the optical and $\gamma$-ray light curves and the SED of the source from the optical to the X-ray bands. \cite{Hudec06} favor this model for 3C 66A and include the source as a candidate to harbor a SBBH. However, the hypothesis of a SBBH to explain all observed quasi-periodic behaviour in blazars can be ruled out according to the recent work by \citet{Sandrinelli18} who compared the prevalence of quasi-periodic behaviour between bright $\gamma$-ray blazars and quasars. \citet{Sandrinelli18} found that the local density of quasi-periodic blazars is much larger than that of 
    quasars. On the other hand, \citet{Sesana18} found that if all quasi-periodic quasars involved a SBBH system, then 
    the expected gravitational wave background at frequencies corresponding to year timescales would be in conflict with 
    that measured with pulsar timing arrays \citep{Foster90}. \cite{Holgado18} similarly concluded that binarity cannot uniquely explain quasi-periodicity in the \textit{Fermi} blazar population on timescales of a few years.
    
    \item Variations in the relativistic jet emission mechanism, either due to geometrical effects such as helical jets or helical structures in the jet \citep{Camenzind92, Rieger04, Mohan15}. The emitting flow moving around an helical path could produce relatively long-term quasi-periodic changes in Doppler boosted flux \citep{Villata99, Ostorero04, Raiteri10}. In this case, the variation is produced by geometrical effects due to changes in the viewing angle or the observation of different emitting regions of the jet at different times. In a recent work \citet{Casadio19} found evidence for the presence of large scale helical magnetic fields from high-resolution mm-VLBI observations of CTA~102. \citet{Raiteri17} could explain a large flare in CTA~102 occurring at the end of 2016 as variations in the orientation of the viewing angle of the jet emitting regions and, hence, their relative Doppler factors. They propose that the orientation of different jet regions may change due to magneto-hydrodynamic (MHD) instabilities \citep{Mignone10} or rotation of a twisted jet. Indeed, jets show large scale inhomogeneities which can be attributed to different mechanisms: re-confinements and relativistic shocks \citep{Marscher85}, Kelvin-Helmholtz instabilities, helical patterns or turbulent cells \citep{Marscher14}. Alternatively, variations in the energy outflow of the jet could be induced by instabilities in the accretion flow \citep{Tchekhovskoy11, Godfrey12}. The innermost part of accretion disks around super-massive black holes is pulsationally unstable. MHD simulations of magnetically choked accretion flows have shown that oscillations with a period of tens of days can be produced in an accretion disk around a spinning black hole of $M_{BH} \simeq 10^9 M_\odot$ \citep{McKinney12}. This period is still relatively short compared to those found in this work. 
    
\end{enumerate}

There may be different mechanisms responsible for the two cases that we have analyzed. 3C 66A is a BL Lac object, which is supposed to contain a radiatively inefficient accretion disk \citep{Fragile09}. A plausible mechanism could be disk instabilities transmitted to the jet \citep{Godfrey12}, but not affecting the disk emission. 
In this scenario, there is no clear reason to expect a correlation between the total and polarized flux, since these instabilities should propagate through the jet without affecting any structure such as magnetic fields that can affect the polarization of the emitted flux. However, this model may be disfavoured by the timescales involved in this scenario, on the order of 10$^{3}$ years for black hole masses of $\sim$10$^{8}$M$_{\odot}$ \citep{Godfrey12}.
On the other hand, the fact that we find quasi-periodic behaviour not only in integrated light, but also its linear polarization and a possible hint at high-energies in B2 1633+38 indicates that the responsible mechanism must be persistent in time. B2 1633+38 is a high luminosity FSRQ blazar, which may develop a powerful jet with stable structures such as helical structures, reconfinements or shock fronts, in a precessing or bent structure. The idea of an helical-like magnetic field is favoured by \cite{hagenthorn2019}, based on the different directions of the polarization vector in different epochs. Moreover, a shock wave propagating along a helical path in the blazar's jet was proposed to explain the multi-frequency behavior of an outburst in 2011 of the blazar S5~0716+71 \citep{larionov2013}. The strong correlation between continuum flux and polarization degree of B2 1633+38 is consistent with the result found by \cite{raiteri2012} using a much shorter time window.
Such behaviour was explained in terms of an inhomogeneous jet in which variability is caused by changes of the emitting region viewing angle, $\theta = 2.6^{o}-5.3^{o}$, coupled with a high Lorentz factor $\Gamma = 31.1$ \citep[see also][]{Raiteri13}. 
Different jet models, helical magnetic fields by \citep{Lyutikov2005} or transverse shock waves \citep{Hughes1985}, predict a maximum value for the polarization degree at the viewing angle, $\theta_{max} \sim 1/\Gamma$. For viewing angles smaller than $\theta_{max}$, the polarization decreases whereas the flux increases when the angle gets closer to zero, due to Doppler boosting. The opposite happens for viewing angles larger than $\theta_{max}$, then a positive correlation is observed between the flux and the polarization degree, as proposed for B2 1633+38.
In the case of 3C 66A, the same model could be applied considering a smaller Lorentz factor, $\Gamma = 16.1$ \citep{celotti2008, jorstad2017} and a similar viewing angle, $\theta = 1.5^{o}-4.5^{o}$ \citep[Fig. 17 in][]{Raiteri13}. In this case, the viewing angle is close to $\theta_{max}$, and no strong correlation is expected between total and polarized optical fluxes. 
Note that determination of Lorentz factor and inclination angles depends largely on the used method: rapid variability observed at high energies and modelling of SEDs usually predict very high values for $\Gamma$ (50-100), whereas high-resolution radio observations (VLBA) imply much lower values for $\Gamma$ \citep[the so called Doppler crisis, see][]{Tavecchio2006, Piner2018}. Furthermore, the determination of $\Gamma$ based on VLBA observations depends on the knotty structure observed in the relativistic jet (e.g. for 3C 66A, \cite{jorstad2017} found $\Gamma\sim30$ and $\Gamma\sim13$ for knots B1 and B2, respectively).

On the other hand, detection of quasi-periodicity on different timescales in the same object may suggest the presence of several inhomogeneities in the jet structure. This behaviour is more difficult to explain with a SBBH model scenario in which a single quasi-periodicity timescale is expected. Moreover, this model predicts quasi-periodicities in timescales $>$10 years \citep[e.g.][]{romero2000}. Shorter timescales explained with a SBBH model would imply very close binary systems. The detection of these systems is very improbable due to their very short lifetime \citep{rieger2019}.

\section{Conclusions}\label{conclusions}
We studied the total and polarized optical flux in V and R bands, along with the $\gamma$-ray emission of the BL Lac object 3C 66A and the FSRQ B2 1633+38, searching for quasi-periodic variability. The optical and polarized light curves extend more than 14 years for 3C 66A and more than 12 years for B2 1633+38, while the $\gamma$-ray light curves span nearly 11 years.
For this purpose, we used 3 different methods to perform the variability analysis such as the ZDCF, the LS periodogram and the WWZ.

Evidence of quasi-periodic variability was found in the optical light curves of 3C 66A with a period of $\sim$3 years at a CL of between 2$\sigma$ and 5$\sigma$ depending on the analysis method. Also, a possible secondary period of $\sim$2.3 years was found in the R band by dividing the light curve into a high state part compatible with the 3 year period, and a low state part that shows this shorter period value, which is compatible with the results presented by \cite{belokon2003} and \cite{kaur2017}. In addition, we identified a period of $\sim$180 days, previously detected by \cite{fan2018}. No periodic behaviour was seen neither in the polarized or $\gamma$-ray light curves.

On the other hand, evidences of quasi-periodicity were detected in both the optical total and polarized flux light curves of B2 1633+38, with an approximate period of $\sim$1.9 years at a CL of between 2$\sigma$ and 5$\sigma$ depending on the variability analysis method used. This appears to be the first report of quasi-periodic variability in optical polarized emission from a blazar and is a consequence of the periodicity detected in the total flux combined with the correlation between flux and the polarization degree pointed out by \cite{raiteri2012}. A hint of periodicity was also detected in the $\gamma$-ray light curve of this source. However, the emission is dominated by flares only during a part of the $\gamma$-ray light curve, combined with a long quiescent state. Thus, we cannot confirm that the recurrence of these flares is periodic, or produced by chance.

Different jet emission mechanisms were proposed to account for the results. While periodicity in 3C 66A could be supported by SBBH models due to the source being radiatively inefficient \citep{Fragile09} and a being classified as a SBBH candidate \citep{Hudec06}, the periodicity in B2 1633+38 may be explained by geometrical effects in the jet such as helical structures or shock fronts \citep{Marscher85, Marscher14}. The same scenario could also be valid in the case of 3C 66A. Based only on the results for two blazars, we obviously cannot generalize the assignment of different mechanisms causing periodic variability behavior to BL Lac objects and FSRQs. Future studies of larger samples of blazars will need to be made to clarify the nature of quasi-periodic variability in blazars.

\section*{Acknowledgements}

JOS thanks the support from grant FPI-SO from the Spanish Ministry of Economy and Competitiveness (MINECO) (research project SEV-2015-0548-17-3 and predoctoral contract BES-2017-082171).

JBG acknowledges the support of the Viera y Clavijo program funded by ACIISI and ULL.

The \textit{Fermi} LAT Collaboration acknowledges generous ongoing support from a number of agencies and institutes that have supported both the development and the operation of the LAT as well as scientific data analysis. These include the National Aeronautics and Space Administration and the Department of Energy in the United States, the Commissariat \`a l'Energie Atomique
and the Centre National de la Recherche Scientifique / Institut National de Physique Nucl\'eaire et de Physique des Particules in France, the Agenzia Spaziale Italiana and the Istituto Nazionale di Fisica Nucleare in Italy, the Ministry of Education, Culture, Sports, Science and Technology (MEXT), High Energy Accelerator Research Organization (KEK) and Japan Aerospace Exploration Agency (JAXA) in Japan, and the K.~A.~Wallenberg Foundation, the Swedish Research Council and the Swedish National Space Board in Sweden. Additional support for science analysis during the operations phase is gratefully acknowledged from the Istituto Nazionale di Astrofisica in Italy and the Centre National d'\'Etudes Spatiales in France. This work performed in part under DOE Contract DE-AC02-76SF00515.

Data from the Steward Observatory spectropolarimetric monitoring project were used. This program is supported by Fermi Guest Investigator grants NNX08AW56G, NNX09AU10G, NNX12AO93G and NNX15AU81G.

The CSS survey is funded by the National Aeronautics and Space
Administration under Grant No. NNG05GF22G issued through the Science Mission Directorate Near-Earth Objects Observations Program. The CRTS survey is supported by the U.S.~National Science Foundation under
grants AST-0909182.

The authors gratefully acknowledge the optical observations provided by the Katzman Automatic Imaging Telescope (KAIT) AGN monitoring program team.

Partly based on data collected by the WEBT collaboration, which are stored in the WEBT archive at the Osservatorio Astrofisico di Torino - INAF (http://www.oato.inaf.it/blazars/webt/); for questions regarding their availability, please contact the WEBT President Massimo Villata ({\tt massimo.villata@inaf.it}).

We acknowledge financial contribution from the agreement ASI-INAF n.2017-14-H.0 and from the contract PRIN-SKA-CTA-INAF 2016.

St.Petersburg University team acknowledges support from Russian Science Foundation grant 17-12-01029.

NCS acknowledge support by the Science and Technology Facilities Council (STFC), and from STFC grant ST/M001326/1.

Based (partly) on data obtained with the STELLA robotic telescopes in Tenerife, an AIP facility jointly operated by AIP and IAC.

This work makes use of observations from the LCOGT network.

The authors thank the anonymous referee for his/her constructive comments and suggestions.




\bibliographystyle{mnras}
\bibliography{biblio} 

\begin{thebibliography}{}
\makeatletter
\relax
\def\mn@urlcharsother{\let\do\@makeother \do\$\do\&\do\#\do\^\do\_\do\%\do\~}
\def\mn@doi{\begingroup\mn@urlcharsother \@ifnextchar [ {\mn@doi@}
  {\mn@doi@[]}}
\def\mn@doi@[#1]#2{\def\@tempa{#1}\ifx\@tempa\@empty \href
  {http://dx.doi.org/#2} {doi:#2}\else \href {http://dx.doi.org/#2} {#1}\fi
  \endgroup}
\def\mn@eprint#1#2{\mn@eprint@#1:#2::\@nil}
\def\mn@eprint@arXiv#1{\href {http://arxiv.org/abs/#1} {{\tt arXiv:#1}}}
\def\mn@eprint@dblp#1{\href {http://dblp.uni-trier.de/rec/bibtex/#1.xml}
  {dblp:#1}}
\def\mn@eprint@#1:#2:#3:#4\@nil{\def\@tempa {#1}\def\@tempb {#2}\def\@tempc
  {#3}\ifx \@tempc \@empty \let \@tempc \@tempb \let \@tempb \@tempa \fi \ifx
  \@tempb \@empty \def\@tempb {arXiv}\fi \@ifundefined
  {mn@eprint@\@tempb}{\@tempb:\@tempc}{\expandafter \expandafter \csname
  mn@eprint@\@tempb\endcsname \expandafter{\@tempc}}}

\bibitem[\protect\citeauthoryear{{Abdo} et~al.,}{{Abdo}
  et~al.}{2011}]{abdo2011}
{Abdo} A.~A.,  et~al., 2011, \mn@doi [\apj] {10.1088/0004-637X/726/1/43}, \href
  {https://ui.adsabs.harvard.edu/abs/2011ApJ...726...43A} {726, 43}

\bibitem[\protect\citeauthoryear{{Acciari} et~al.,}{{Acciari}
  et~al.}{2009}]{acciari2009}
{Acciari} V.~A.,  et~al., 2009, \mn@doi [\apjl] {10.1088/0004-637X/693/2/L104},
  \href {https://ui.adsabs.harvard.edu/abs/2009ApJ...693L.104A} {693, L104}

\bibitem[\protect\citeauthoryear{{Ackermann} et~al.,}{{Ackermann}
  et~al.}{2015a}]{ackermann2015}
{Ackermann} M.,  et~al., 2015a, \mn@doi [\apj] {10.1088/0004-637X/810/1/14},
  \href {https://ui.adsabs.harvard.edu/abs/2015ApJ...810...14A} {810, 14}

\bibitem[\protect\citeauthoryear{{Ackermann} et~al.,}{{Ackermann}
  et~al.}{2015b}]{Ackermann15}
{Ackermann} M.,  et~al., 2015b, \mn@doi [\apjl] {10.1088/2041-8205/813/2/L41},
  \href {https://ui.adsabs.harvard.edu/abs/2015ApJ...813L..41A} {813, L41}

\bibitem[\protect\citeauthoryear{{Ackermann} et~al.,}{{Ackermann}
  et~al.}{2016}]{ackermann2016}
{Ackermann} M.,  et~al., 2016, \mn@doi [\apjs] {10.3847/0067-0049/222/1/5},
  \href {https://ui.adsabs.harvard.edu/abs/2016ApJS..222....5A} {222, 5}

\bibitem[\protect\citeauthoryear{{Aleksi{\'c}} et~al.,}{{Aleksi{\'c}}
  et~al.}{2011}]{aleksic2011}
{Aleksi{\'c}} J.,  et~al., 2011, \mn@doi [\apj] {10.1088/0004-637X/726/2/58},
  \href {https://ui.adsabs.harvard.edu/abs/2011ApJ...726...58A} {726, 58}

\bibitem[\protect\citeauthoryear{{Alexander}}{{Alexander}}{1997}]{zdcf2}
{Alexander} T.,  1997, {Is AGN Variability Correlated with Other AGN
  Properties? ZDCF Analysis of Small Samples of Sparse Light Curves}.
p.~163, \mn@doi{10.1007/978-94-015-8941-3_14}

\bibitem[\protect\citeauthoryear{{Alexander}}{{Alexander}}{2013}]{zdcf}
{Alexander} T.,  2013, arXiv e-prints, \href
  {https://ui.adsabs.harvard.edu/abs/2013arXiv1302.1508A} {p. arXiv:1302.1508}

\bibitem[\protect\citeauthoryear{{An}, {Lu}  \& {Wang}}{{An}
  et~al.}{2016}]{wwz_ex}
{An} T.,  {Lu} X.-L.,   {Wang} J.-Y.,  2016, \mn@doi [\aap]
  {10.1051/0004-6361/201526182}, \href
  {https://ui.adsabs.harvard.edu/abs/2016A&A...585A..89A} {585, A89}

\bibitem[\protect\citeauthoryear{{Atwood} et~al.,}{{Atwood}
  et~al.}{2009}]{Fermitelescope}
{Atwood} W.~B.,  et~al., 2009, \mn@doi [\apj] {10.1088/0004-637X/697/2/1071},
  \href {https://ui.adsabs.harvard.edu/abs/2009ApJ...697.1071A} {697, 1071}

\bibitem[\protect\citeauthoryear{{Barbieri}, {Romano}, {di Serego}  \&
  {Zambon}}{{Barbieri} et~al.}{1977}]{barbieri1977}
{Barbieri} C.,  {Romano} G.,  {di Serego} S.,   {Zambon} M.,  1977, \aap, \href
  {https://ui.adsabs.harvard.edu/abs/1977A&A....59..419B} {59, 419}

\bibitem[\protect\citeauthoryear{{Belokon} \& {Babadzhanyants}}{{Belokon} \&
  {Babadzhanyants}}{2003}]{belokon2003}
{Belokon} E.~T.,  {Babadzhanyants} M.~K.,  2003, in High Energy Blazar
  Astronomy. p.~205

\bibitem[\protect\citeauthoryear{{Bhatta} et~al.,}{{Bhatta}
  et~al.}{2016}]{wwzex}
{Bhatta} G.,  et~al., 2016, \mn@doi [\apj] {10.3847/0004-637X/832/1/47}, \href
  {https://ui.adsabs.harvard.edu/abs/2016ApJ...832...47B} {832, 47}

\bibitem[\protect\citeauthoryear{{Bonning} et~al.,}{{Bonning}
  et~al.}{2009}]{Bonning09}
{Bonning} E.~W.,  et~al., 2009, \mn@doi [\apjl] {10.1088/0004-637X/697/2/L81},
  \href {https://ui.adsabs.harvard.edu/abs/2009ApJ...697L..81B} {697, L81}

\bibitem[\protect\citeauthoryear{{B{\"o}ttcher}}{{B{\"o}ttcher}}{2019}]{bottcher19}
{B{\"o}ttcher} M.,  2019, \mn@doi [Galaxies] {10.3390/galaxies7010020}, \href
  {https://ui.adsabs.harvard.edu/abs/2019Galax...7...20B} {7, 20}

\bibitem[\protect\citeauthoryear{{B{\"o}ttcher} et~al.,}{{B{\"o}ttcher}
  et~al.}{2005}]{bottcher2005}
{B{\"o}ttcher} M.,  et~al., 2005, \mn@doi [\apj] {10.1086/432609}, \href
  {https://ui.adsabs.harvard.edu/abs/2005ApJ...631..169B} {631, 169}

\bibitem[\protect\citeauthoryear{{B{\"o}ttcher} et~al.,}{{B{\"o}ttcher}
  et~al.}{2009}]{bottcher2009}
{B{\"o}ttcher} M.,  et~al., 2009, \mn@doi [\apj] {10.1088/0004-637X/694/1/174},
  \href {https://ui.adsabs.harvard.edu/abs/2009ApJ...694..174B} {694, 174}

\bibitem[\protect\citeauthoryear{{Camenzind} \& {Krockenberger}}{{Camenzind} \&
  {Krockenberger}}{1992}]{Camenzind92}
{Camenzind} M.,  {Krockenberger} M.,  1992, \aap, \href
  {https://ui.adsabs.harvard.edu/abs/1992A&A...255...59C} {255, 59}

\bibitem[\protect\citeauthoryear{{Caproni}, {Abraham}, {Motter}  \&
  {Monteiro}}{{Caproni} et~al.}{2017}]{Caproni17}
{Caproni} A.,  {Abraham} Z.,  {Motter} J.~C.,   {Monteiro} H.,  2017, \mn@doi
  [\apjl] {10.3847/2041-8213/aa9fea}, \href
  {https://ui.adsabs.harvard.edu/abs/2017ApJ...851L..39C} {851, L39}

\bibitem[\protect\citeauthoryear{{Carnerero} et~al.,}{{Carnerero}
  et~al.}{2015}]{Carnerero15}
{Carnerero} M.~I.,  et~al., 2015, \mn@doi [\mnras] {10.1093/mnras/stv823},
  \href {https://ui.adsabs.harvard.edu/abs/2015MNRAS.450.2677C} {450, 2677}

\bibitem[\protect\citeauthoryear{{Casadio} et~al.,}{{Casadio}
  et~al.}{2019}]{Casadio19}
{Casadio} C.,  et~al., 2019, \mn@doi [\aap] {10.1051/0004-6361/201834519},
  \href {https://ui.adsabs.harvard.edu/abs/2019A&A...622A.158C} {622, A158}

\bibitem[\protect\citeauthoryear{{Cavaliere}, {Tavani}  \&
  {Vittorini}}{{Cavaliere} et~al.}{2017}]{Cavaliere17}
{Cavaliere} A.,  {Tavani} M.,   {Vittorini} V.,  2017, \mn@doi [\apj]
  {10.3847/1538-4357/836/2/220}, \href
  {https://ui.adsabs.harvard.edu/abs/2017ApJ...836..220C} {836, 220}

\bibitem[\protect\citeauthoryear{{Celotti} \& {Ghisellini}}{{Celotti} \&
  {Ghisellini}}{2008}]{celotti2008}
{Celotti} A.,  {Ghisellini} G.,  2008, \mn@doi [\mnras]
  {10.1111/j.1365-2966.2007.12758.x}, \href
  {https://ui.adsabs.harvard.edu/abs/2008MNRAS.385..283C} {385, 283}

\bibitem[\protect\citeauthoryear{{Chatterjee} et~al.,}{{Chatterjee}
  et~al.}{2008}]{zdcfsignej}
{Chatterjee} R.,  et~al., 2008, \mn@doi [\apj] {10.1086/592598}, \href
  {https://ui.adsabs.harvard.edu/abs/2008ApJ...689...79C} {689, 79}

\bibitem[\protect\citeauthoryear{{Chatterjee} et~al.,}{{Chatterjee}
  et~al.}{2012}]{Chatterjee12}
{Chatterjee} R.,  et~al., 2012, \mn@doi [\apj] {10.1088/0004-637X/749/2/191},
  \href {https://ui.adsabs.harvard.edu/abs/2012ApJ...749..191C} {749, 191}

\bibitem[\protect\citeauthoryear{Cleveland}{Cleveland}{1979}]{cleveland1979}
Cleveland W.~S.,  1979, Journal of the American statistical association, 74,
  829

\bibitem[\protect\citeauthoryear{{Covino}, {Sandrinelli}  \& {Treves}}{{Covino}
  et~al.}{2019}]{Covino2019}
{Covino} S.,  {Sandrinelli} A.,   {Treves} A.,  2019, \mn@doi [\mnras]
  {10.1093/mnras/sty2720}, \href
  {https://ui.adsabs.harvard.edu/abs/2019MNRAS.482.1270C} {482, 1270}

\bibitem[\protect\citeauthoryear{{Drake} et~al.,}{{Drake}
  et~al.}{2009}]{catalina_survey}
{Drake} A.~J.,  et~al., 2009, \mn@doi [\apj] {10.1088/0004-637X/696/1/870},
  \href {https://ui.adsabs.harvard.edu/abs/2009ApJ...696..870D} {696, 870}

\bibitem[\protect\citeauthoryear{{Edelson} \& {Krolik}}{{Edelson} \&
  {Krolik}}{1989}]{dcf}
{Edelson} R.~A.,  {Krolik} J.~H.,  1989, in {Osterbrock} D.~E.,  {Miller}
  J.~S.,  eds,  IAU Symposium Vol. 134, Active Galactic Nuclei. p.~96

\bibitem[\protect\citeauthoryear{{Edelson} et~al.,}{{Edelson}
  et~al.}{1995}]{zdcfsignej2}
{Edelson} R.,  et~al., 1995, \mn@doi [\apj] {10.1086/175059}, \href
  {https://ui.adsabs.harvard.edu/abs/1995ApJ...438..120E} {438, 120}

\bibitem[\protect\citeauthoryear{{Emmanoulopoulos}, {McHardy}  \&
  {Papadakis}}{{Emmanoulopoulos} et~al.}{2013}]{simLC}
{Emmanoulopoulos} D.,  {McHardy} I.~M.,   {Papadakis} I.~E.,  2013, \mn@doi
  [\mnras] {10.1093/mnras/stt764}, \href
  {https://ui.adsabs.harvard.edu/abs/2013MNRAS.433..907E} {433, 907}

\bibitem[\protect\citeauthoryear{{Fan} et~al.,}{{Fan} et~al.}{2006}]{fan2006}
{Fan} J.~H.,  et~al., 2006, Chinese Journal of Astronomy and Astrophysics
  Supplement, \href {https://ui.adsabs.harvard.edu/abs/2006ChJAS...6b.333F} {6,
  333}

\bibitem[\protect\citeauthoryear{{Fan} et~al.,}{{Fan} et~al.}{2018}]{fan2018}
{Fan} J.~H.,  et~al., 2018, \mn@doi [\aj] {10.3847/1538-3881/aaa547}, \href
  {https://ui.adsabs.harvard.edu/abs/2018AJ....155...90F} {155, 90}

\bibitem[\protect\citeauthoryear{{Finke}}{{Finke}}{2013}]{finke2013}
{Finke} J.~D.,  2013, \mn@doi [\apj] {10.1088/0004-637X/763/2/134}, \href
  {https://ui.adsabs.harvard.edu/abs/2013ApJ...763..134F} {763, 134}

\bibitem[\protect\citeauthoryear{{Finke}, {Shields}, {B{\"o}ttcher}  \&
  {Basu}}{{Finke} et~al.}{2008}]{finke2008}
{Finke} J.~D.,  {Shields} J.~C.,  {B{\"o}ttcher} M.,   {Basu} S.,  2008,
  \mn@doi [\aap] {10.1051/0004-6361:20078492}, \href
  {https://ui.adsabs.harvard.edu/abs/2008A&A...477..513F} {477, 513}

\bibitem[\protect\citeauthoryear{{Foster}}{{Foster}}{1996}]{wwz}
{Foster} G.,  1996, \mn@doi [\aj] {10.1086/118137}, \href
  {https://ui.adsabs.harvard.edu/abs/1996AJ....112.1709F} {112, 1709}

\bibitem[\protect\citeauthoryear{{Foster} \& {Backer}}{{Foster} \&
  {Backer}}{1990}]{Foster90}
{Foster} R.~S.,  {Backer} D.~C.,  1990, \mn@doi [\apj] {10.1086/169195}, \href
  {https://ui.adsabs.harvard.edu/abs/1990ApJ...361..300F} {361, 300}

\bibitem[\protect\citeauthoryear{{Fragile} \& {Meier}}{{Fragile} \&
  {Meier}}{2009}]{Fragile09}
{Fragile} P.~C.,  {Meier} D.~L.,  2009, \mn@doi [\apj]
  {10.1088/0004-637X/693/1/771}, \href
  {https://ui.adsabs.harvard.edu/abs/2009ApJ...693..771F} {693, 771}

\bibitem[\protect\citeauthoryear{{Furniss}, {Fumagalli}, {Danforth}, {Williams}
   \& {Prochaska}}{{Furniss} et~al.}{2013}]{furniss2013}
{Furniss} A.,  {Fumagalli} M.,  {Danforth} C.,  {Williams} D.~A.,   {Prochaska}
  J.~X.,  2013, \mn@doi [\apj] {10.1088/0004-637X/766/1/35}, \href
  {https://ui.adsabs.harvard.edu/abs/2013ApJ...766...35F} {766, 35}

\bibitem[\protect\citeauthoryear{{Ghosh} \& {Soundararajaperumal}}{{Ghosh} \&
  {Soundararajaperumal}}{1995}]{ghosh1995}
{Ghosh} K.~K.,  {Soundararajaperumal} S.,  1995, \mn@doi [\apjs]
  {10.1086/192207}, \href
  {https://ui.adsabs.harvard.edu/abs/1995ApJS..100...37G} {100, 37}

\bibitem[\protect\citeauthoryear{{Godfrey} et~al.,}{{Godfrey}
  et~al.}{2012}]{Godfrey12}
{Godfrey} L.~E.~H.,  et~al., 2012, \mn@doi [\apjl]
  {10.1088/2041-8205/758/2/L27}, \href
  {https://ui.adsabs.harvard.edu/abs/2012ApJ...758L..27G} {758, L27}

\bibitem[\protect\citeauthoryear{{Graham} et~al.,}{{Graham}
  et~al.}{2015}]{Graham15}
{Graham} M.~J.,  et~al., 2015, \mn@doi [\nat] {10.1038/nature14143}, \href
  {https://ui.adsabs.harvard.edu/abs/2015Natur.518...74G} {518, 74}

\bibitem[\protect\citeauthoryear{{Grossmann}, {Kronland-Martinet}  \&
  {Morlet}}{{Grossmann} et~al.}{1989}]{morlet_wavelet}
{Grossmann} A.,  {Kronland-Martinet} R.,   {Morlet} J.,  1989, in {Combes}
  J.-M.,  {Grossmann} A.,   {Tchamitchian} P.,  eds, Wavelets. Time-Frequency
  Methods and Phase Space. p.~2

\bibitem[\protect\citeauthoryear{{Gupta}}{{Gupta}}{2018}]{gupta2018}
{Gupta} A.,  2018, \mn@doi [Galaxies] {10.3390/galaxies6010001}, \href
  {https://ui.adsabs.harvard.edu/abs/2018Galax...6....1G} {6, 1}

\bibitem[\protect\citeauthoryear{{Hagen-Thorn} et~al.,}{{Hagen-Thorn}
  et~al.}{2019}]{hagenthorn2019}
{Hagen-Thorn} V.~A.,  et~al., 2019, \mn@doi [Astronomy Reports]
  {10.1134/S1063772919040036}, \href
  {https://ui.adsabs.harvard.edu/abs/2019ARep...63..378H} {63, 378}

\bibitem[\protect\citeauthoryear{Han, Wang, Lin  \& An}{Han
  et~al.}{2012}]{wavelet2}
Han X.,  Wang J.,  Lin J.,   An T.,  2012, in 2012 8th International Conference
  on Wireless Communications, Networking and Mobile Computing. pp~1--4

\bibitem[\protect\citeauthoryear{{Holgado}, {Sesana}, {Sandrinelli}, {Covino},
  {Treves}, {Liu}  \& {Ricker}}{{Holgado} et~al.}{2018}]{Holgado18}
{Holgado} A.~M.,  {Sesana} A.,  {Sandrinelli} A.,  {Covino} S.,  {Treves} A.,
  {Liu} X.,   {Ricker} P.,  2018, \mn@doi [\mnras] {10.1093/mnrasl/sly158},
  \href {https://ui.adsabs.harvard.edu/abs/2018MNRAS.481L..74H} {481, L74}

\bibitem[\protect\citeauthoryear{{Hovatta}, {Tornikoski}, {Lainela}, {Lehto},
  {Valtaoja}, {Torniainen}, {Aller}  \& {Aller}}{{Hovatta}
  et~al.}{2007}]{hovatta2007}
{Hovatta} T.,  {Tornikoski} M.,  {Lainela} M.,  {Lehto} H.~J.,  {Valtaoja} E.,
  {Torniainen} I.,  {Aller} M.~F.,   {Aller} H.~D.,  2007, \mn@doi [\aap]
  {10.1051/0004-6361:20077529}, \href
  {https://ui.adsabs.harvard.edu/abs/2007A&A...469..899H} {469, 899}

\bibitem[\protect\citeauthoryear{{Hudec} \& {Basta}}{{Hudec} \&
  {Basta}}{2006}]{Hudec06}
{Hudec} R.,  {Basta} M.,  2006, \mn@doi [Chinese Journal of Astronomy and
  Astrophysics Supplement] {10.1088/1009-9271/6/S1/32}, \href
  {https://ui.adsabs.harvard.edu/abs/2006ChJAS...6a.253H} {6, 253}

\bibitem[\protect\citeauthoryear{{Hughes}, {Aller}  \& {Aller}}{{Hughes}
  et~al.}{1985}]{Hughes1985}
{Hughes} P.~A.,  {Aller} H.~D.,   {Aller} M.~F.,  1985, \mn@doi [\apj]
  {10.1086/163611}, \href
  {https://ui.adsabs.harvard.edu/abs/1985ApJ...298..301H} {298, 301}

\bibitem[\protect\citeauthoryear{{Jorstad} et~al.,}{{Jorstad}
  et~al.}{2017}]{jorstad2017}
{Jorstad} S.~G.,  et~al., 2017, \mn@doi [\apj] {10.3847/1538-4357/aa8407},
  \href {https://ui.adsabs.harvard.edu/abs/2017ApJ...846...98J} {846, 98}

\bibitem[\protect\citeauthoryear{{Kaur}, {Sameer}, {Baliyan}  \&
  {Ganesh}}{{Kaur} et~al.}{2017}]{kaur2017}
{Kaur} N.,  {Sameer} {Baliyan} K.~S.,   {Ganesh} S.,  2017, \mn@doi [\mnras]
  {10.1093/mnras/stx965}, \href
  {https://ui.adsabs.harvard.edu/abs/2017MNRAS.469.2305K} {469, 2305}

\bibitem[\protect\citeauthoryear{{Lainela} et~al.,}{{Lainela}
  et~al.}{1999}]{lainela1999}
{Lainela} M.,  et~al., 1999, \mn@doi [\apj] {10.1086/307599}, \href
  {https://ui.adsabs.harvard.edu/abs/1999ApJ...521..561L} {521, 561}

\bibitem[\protect\citeauthoryear{{Larionov} et~al.,}{{Larionov}
  et~al.}{2013}]{larionov2013}
{Larionov} V.~M.,  et~al., 2013, \mn@doi [\apj] {10.1088/0004-637X/768/1/40},
  \href {https://ui.adsabs.harvard.edu/abs/2013ApJ...768...40L} {768, 40}

\bibitem[\protect\citeauthoryear{{Lehto} \& {Valtonen}}{{Lehto} \&
  {Valtonen}}{1996}]{lehto1996}
{Lehto} H.~J.,  {Valtonen} M.~J.,  1996, \mn@doi [\apj] {10.1086/176962}, \href
  {https://ui.adsabs.harvard.edu/abs/1996ApJ...460..207L} {460, 207}

\bibitem[\protect\citeauthoryear{{Li}, {Filippenko}, {Chornock}  \& {Jha}}{{Li}
  et~al.}{2003}]{kait}
{Li} W.,  {Filippenko} A.~V.,  {Chornock} R.,   {Jha} S.,  2003, \mn@doi
  [\pasp] {10.1086/376432}, \href
  {https://ui.adsabs.harvard.edu/abs/2003PASP..115..844L} {115, 844}

\bibitem[\protect\citeauthoryear{{Liao}, {Bai}, {Liu}, {Weng}, {Chen}  \&
  {Li}}{{Liao} et~al.}{2014}]{Liao14}
{Liao} N.~H.,  {Bai} J.~M.,  {Liu} H.~T.,  {Weng} S.~S.,  {Chen} L.,   {Li} F.,
   2014, \mn@doi [\apj] {10.1088/0004-637X/783/2/83}, \href
  {https://ui.adsabs.harvard.edu/abs/2014ApJ...783...83L} {783, 83}

\bibitem[\protect\citeauthoryear{{Liu}, {Jiang}, {Shen}  \& {Karouzos}}{{Liu}
  et~al.}{2010}]{liu2010}
{Liu} Y.,  {Jiang} D.~R.,  {Shen} Z.~Q.,   {Karouzos} M.,  2010, \mn@doi [\aap]
  {10.1051/0004-6361/201014113}, \href
  {https://ui.adsabs.harvard.edu/abs/2010A&A...522A...5L} {522, A5}

\bibitem[\protect\citeauthoryear{{Liu} et~al.,}{{Liu} et~al.}{2017}]{liu2017}
{Liu} X.,  et~al., 2017, \mn@doi [\mnras] {10.1093/mnras/stx1062}, \href
  {https://ui.adsabs.harvard.edu/abs/2017MNRAS.469.2457L} {469, 2457}

\bibitem[\protect\citeauthoryear{{Lomb}}{{Lomb}}{1976}]{lomb1976}
{Lomb} N.~R.,  1976, \mn@doi [\apss] {10.1007/BF00648343}, \href
  {https://ui.adsabs.harvard.edu/abs/1976Ap&SS..39..447L} {39, 447}

\bibitem[\protect\citeauthoryear{{Lyutikov}, {Pariev}  \& {Gabuzda}}{{Lyutikov}
  et~al.}{2005}]{Lyutikov2005}
{Lyutikov} M.,  {Pariev} V.~I.,   {Gabuzda} D.~C.,  2005, \mn@doi [\mnras]
  {10.1111/j.1365-2966.2005.08954.x}, \href
  {https://ui.adsabs.harvard.edu/abs/2005MNRAS.360..869L} {360, 869}

\bibitem[\protect\citeauthoryear{{Marscher}}{{Marscher}}{2014}]{Marscher14}
{Marscher} A.~P.,  2014, \mn@doi [\apj] {10.1088/0004-637X/780/1/87}, \href
  {https://ui.adsabs.harvard.edu/abs/2014ApJ...780...87M} {780, 87}

\bibitem[\protect\citeauthoryear{{Marscher} \& {Gear}}{{Marscher} \&
  {Gear}}{1985}]{Marscher85}
{Marscher} A.~P.,  {Gear} W.~K.,  1985, \mn@doi [\apj] {10.1086/163592}, \href
  {https://ui.adsabs.harvard.edu/abs/1985ApJ...298..114M} {298, 114}

\bibitem[\protect\citeauthoryear{{Max-Moerbeck}, {Richards}, {Hovatta},
  {Pavlidou}, {Pearson}  \& {Readhead}}{{Max-Moerbeck} et~al.}{2014}]{zdcfsign}
{Max-Moerbeck} W.,  {Richards} J.~L.,  {Hovatta} T.,  {Pavlidou} V.,  {Pearson}
  T.~J.,   {Readhead} A.~C.~S.,  2014, \mn@doi [\mnras]
  {10.1093/mnras/stu1707}, \href
  {https://ui.adsabs.harvard.edu/abs/2014MNRAS.445..437M} {445, 437}

\bibitem[\protect\citeauthoryear{{McKinney}, {Tchekhovskoy}  \& {Bland
  ford}}{{McKinney} et~al.}{2012}]{McKinney12}
{McKinney} J.~C.,  {Tchekhovskoy} A.,   {Bland ford} R.~D.,  2012, \mn@doi
  [\mnras] {10.1111/j.1365-2966.2012.21074.x}, \href
  {https://ui.adsabs.harvard.edu/abs/2012MNRAS.423.3083M} {423, 3083}

\bibitem[\protect\citeauthoryear{{McQuillan}, {Aigrain}  \&
  {Mazeh}}{{McQuillan} et~al.}{2013}]{keplerrotation}
{McQuillan} A.,  {Aigrain} S.,   {Mazeh} T.,  2013, \mn@doi [\mnras]
  {10.1093/mnras/stt536}, \href
  {https://ui.adsabs.harvard.edu/abs/2013MNRAS.432.1203M} {432, 1203}

\bibitem[\protect\citeauthoryear{{Mignone}, {Rossi}, {Bodo}, {Ferrari}  \&
  {Massaglia}}{{Mignone} et~al.}{2010}]{Mignone10}
{Mignone} A.,  {Rossi} P.,  {Bodo} G.,  {Ferrari} A.,   {Massaglia} S.,  2010,
  \mn@doi [\mnras] {10.1111/j.1365-2966.2009.15642.x}, \href
  {https://ui.adsabs.harvard.edu/abs/2010MNRAS.402....7M} {402, 7}

\bibitem[\protect\citeauthoryear{{Miller}, {French}  \& {Hawley}}{{Miller}
  et~al.}{1978}]{miller1978}
{Miller} J.~S.,  {French} H.~B.,   {Hawley} S.~A.,  1978, in {Wolfe} A.~M.,
  ed., BL Lac Objects. pp 176--187

\bibitem[\protect\citeauthoryear{{Mohan} \& {Mangalam}}{{Mohan} \&
  {Mangalam}}{2015}]{Mohan15}
{Mohan} P.,  {Mangalam} A.,  2015, \mn@doi [\apj] {10.1088/0004-637X/805/2/91},
  \href {https://ui.adsabs.harvard.edu/abs/2015ApJ...805...91M} {805, 91}

\bibitem[\protect\citeauthoryear{{Ostorero}, {Villata}  \&
  {Raiteri}}{{Ostorero} et~al.}{2004}]{Ostorero04}
{Ostorero} L.,  {Villata} M.,   {Raiteri} C.~M.,  2004, \mn@doi [\aap]
  {10.1051/0004-6361:20035813}, \href
  {https://ui.adsabs.harvard.edu/abs/2004A&A...419..913O} {419, 913}

\bibitem[\protect\citeauthoryear{{Paiano}, {Landoni}, {Falomo}, {Treves},
  {Scarpa}  \& {Righi}}{{Paiano} et~al.}{2017}]{Paiano17}
{Paiano} S.,  {Landoni} M.,  {Falomo} R.,  {Treves} A.,  {Scarpa} R.,   {Righi}
  C.,  2017, \mn@doi [\apj] {10.3847/1538-4357/837/2/144}, \href
  {https://ui.adsabs.harvard.edu/abs/2017ApJ...837..144P} {837, 144}

\bibitem[\protect\citeauthoryear{{P{\^a}ris} et~al.,}{{P{\^a}ris}
  et~al.}{2017}]{paris2017}
{P{\^a}ris} I.,  et~al., 2017, \mn@doi [\aap] {10.1051/0004-6361/201527999},
  \href {https://ui.adsabs.harvard.edu/abs/2017A&A...597A..79P} {597, A79}

\bibitem[\protect\citeauthoryear{{Piner} \& {Edwards}}{{Piner} \&
  {Edwards}}{2018}]{Piner2018}
{Piner} B.~G.,  {Edwards} P.~G.,  2018, in Fourteenth Marcel Grossmann Meeting
  - MG14. pp 3074--3079, \mn@doi{10.1142/9789813226609_0389}

\bibitem[\protect\citeauthoryear{{Raiteri} et~al.,}{{Raiteri}
  et~al.}{2010}]{Raiteri10}
{Raiteri} C.~M.,  et~al., 2010, \mn@doi [\aap] {10.1051/0004-6361/201015191},
  \href {https://ui.adsabs.harvard.edu/abs/2010A&A...524A..43R} {524, A43}

\bibitem[\protect\citeauthoryear{{Raiteri} et~al.,}{{Raiteri}
  et~al.}{2012}]{raiteri2012}
{Raiteri} C.~M.,  et~al., 2012, \mn@doi [\aap] {10.1051/0004-6361/201219492},
  \href {https://ui.adsabs.harvard.edu/abs/2012A&A...545A..48R} {545, A48}

\bibitem[\protect\citeauthoryear{{Raiteri} et~al.,}{{Raiteri}
  et~al.}{2013}]{Raiteri13}
{Raiteri} C.~M.,  et~al., 2013, \mn@doi [\mnras] {10.1093/mnras/stt1672}, \href
  {https://ui.adsabs.harvard.edu/abs/2013MNRAS.436.1530R} {436, 1530}

\bibitem[\protect\citeauthoryear{{Raiteri} et~al.,}{{Raiteri}
  et~al.}{2017}]{Raiteri17}
{Raiteri} C.~M.,  et~al., 2017, \mn@doi [\nat] {10.1038/nature24623}, \href
  {https://ui.adsabs.harvard.edu/abs/2017Natur.552..374R} {552, 374}

\bibitem[\protect\citeauthoryear{{Rajput}, {Stalin}, {Sahayanathan}, {Rakshit}
  \& {Mandal}}{{Rajput} et~al.}{2019}]{Rajput19}
{Rajput} B.,  {Stalin} C.~S.,  {Sahayanathan} S.,  {Rakshit} S.,   {Mandal}
  A.~K.,  2019, \mn@doi [\mnras] {10.1093/mnras/stz941}, \href
  {https://ui.adsabs.harvard.edu/abs/2019MNRAS.486.1781R} {486, 1781}

\bibitem[\protect\citeauthoryear{{Rieger}}{{Rieger}}{2004}]{Rieger04}
{Rieger} F.~M.,  2004, \mn@doi [\apjl] {10.1086/426018}, \href
  {https://ui.adsabs.harvard.edu/abs/2004ApJ...615L...5R} {615, L5}

\bibitem[\protect\citeauthoryear{{Rieger}}{{Rieger}}{2019}]{rieger2019}
{Rieger} F.,  2019, \mn@doi [Galaxies] {10.3390/galaxies7010028}, \href
  {https://ui.adsabs.harvard.edu/abs/2019Galax...7...28R} {7, 28}

\bibitem[\protect\citeauthoryear{{Romero}, {Chajet}, {Abraham}  \&
  {Fan}}{{Romero} et~al.}{2000}]{romero2000}
{Romero} G.~E.,  {Chajet} L.,  {Abraham} Z.,   {Fan} J.~H.,  2000, \aap, \href
  {https://ui.adsabs.harvard.edu/abs/2000A&A...360...57R} {360, 57}

\bibitem[\protect\citeauthoryear{{Romero}, {Boettcher}, {Markoff}  \&
  {Tavecchio}}{{Romero} et~al.}{2017}]{romero2017}
{Romero} G.~E.,  {Boettcher} M.,  {Markoff} S.,   {Tavecchio} F.,  2017,
  \mn@doi [\ssr] {10.1007/s11214-016-0328-2}, \href
  {https://ui.adsabs.harvard.edu/abs/2017SSRv..207....5R} {207, 5}

\bibitem[\protect\citeauthoryear{{Sandrinelli}, {Covino}, {Dotti}  \&
  {Treves}}{{Sandrinelli} et~al.}{2016}]{sandrinelli2016}
{Sandrinelli} A.,  {Covino} S.,  {Dotti} M.,   {Treves} A.,  2016, \mn@doi
  [\aj] {10.3847/0004-6256/151/3/54}, \href
  {https://ui.adsabs.harvard.edu/abs/2016AJ....151...54S} {151, 54}

\bibitem[\protect\citeauthoryear{{Sandrinelli}, {Covino}, {Treves}, {Holgado},
  {Sesana}, {Lindfors}  \& {Ramazani}}{{Sandrinelli}
  et~al.}{2018}]{Sandrinelli18}
{Sandrinelli} A.,  {Covino} S.,  {Treves} A.,  {Holgado} A.~M.,  {Sesana} A.,
  {Lindfors} E.,   {Ramazani} V.~F.,  2018, \mn@doi [\aap]
  {10.1051/0004-6361/201732550}, \href
  {https://ui.adsabs.harvard.edu/abs/2018A&A...615A.118S} {615, A118}

\bibitem[\protect\citeauthoryear{{Scargle}}{{Scargle}}{1982}]{scargle1982}
{Scargle} J.~D.,  1982, \mn@doi [\apj] {10.1086/160554}, \href
  {https://ui.adsabs.harvard.edu/abs/1982ApJ...263..835S} {263, 835}

\bibitem[\protect\citeauthoryear{{Sesana}, {Haiman}, {Kocsis}  \&
  {Kelley}}{{Sesana} et~al.}{2018}]{Sesana18}
{Sesana} A.,  {Haiman} Z.,  {Kocsis} B.,   {Kelley} L.~Z.,  2018, \mn@doi
  [\apj] {10.3847/1538-4357/aaad0f}, \href
  {https://ui.adsabs.harvard.edu/abs/2018ApJ...856...42S} {856, 42}

\bibitem[\protect\citeauthoryear{{Sillanpaa}, {Haarala}, {Valtonen},
  {Sundelius}  \& {Byrd}}{{Sillanpaa} et~al.}{1988}]{Sillanpaa88}
{Sillanpaa} A.,  {Haarala} S.,  {Valtonen} M.~J.,  {Sundelius} B.,   {Byrd}
  G.~G.,  1988, \mn@doi [\apj] {10.1086/166033}, \href
  {https://ui.adsabs.harvard.edu/abs/1988ApJ...325..628S} {325, 628}

\bibitem[\protect\citeauthoryear{{Smith}, {Montiel}, {Rightley}, {Turner},
  {Schmidt}  \& {Jannuzi}}{{Smith} et~al.}{2009}]{paul_smith}
{Smith} P.~S.,  {Montiel} E.,  {Rightley} S.,  {Turner} J.,  {Schmidt} G.~D.,
  {Jannuzi} B.~T.,  2009, arXiv e-prints, \href
  {https://ui.adsabs.harvard.edu/abs/2009arXiv0912.3621S} {p. arXiv:0912.3621}

\bibitem[\protect\citeauthoryear{{Sobacchi}, {Sormani}  \&
  {Stamerra}}{{Sobacchi} et~al.}{2017}]{Sobacchi17}
{Sobacchi} E.,  {Sormani} M.~C.,   {Stamerra} A.,  2017, \mn@doi [\mnras]
  {10.1093/mnras/stw2684}, \href
  {https://ui.adsabs.harvard.edu/abs/2017MNRAS.465..161S} {465, 161}

\bibitem[\protect\citeauthoryear{{Strassmeier} et~al.,}{{Strassmeier}
  et~al.}{2004}]{stella}
{Strassmeier} K.~G.,  et~al., 2004, \mn@doi [Astronomische Nachrichten]
  {10.1002/asna.200410273}, \href
  {https://ui.adsabs.harvard.edu/abs/2004AN....325..527S} {325, 527}

\bibitem[\protect\citeauthoryear{{Tavecchio}}{{Tavecchio}}{2006}]{Tavecchio2006}
{Tavecchio} F.,  2006, in The Tenth Marcel Grossmann Meeting. On recent
  developments in theoretical and experimental general relativity, gravitation
  and relativistic field theories. p.~512 (\mn@eprint {arXiv}
  {astro-ph/0401590}), \mn@doi{10.1142/9789812704030_0031}

\bibitem[\protect\citeauthoryear{{Tchekhovskoy}, {Narayan}  \&
  {McKinney}}{{Tchekhovskoy} et~al.}{2011}]{Tchekhovskoy11}
{Tchekhovskoy} A.,  {Narayan} R.,   {McKinney} J.~C.,  2011, \mn@doi [\mnras]
  {10.1111/j.1745-3933.2011.01147.x}, \href
  {https://ui.adsabs.harvard.edu/abs/2011MNRAS.418L..79T} {418, L79}

\bibitem[\protect\citeauthoryear{{The Fermi-LAT collaboration}}{{The Fermi-LAT
  collaboration}}{2019}]{fermi4fgl}
{The Fermi-LAT collaboration} 2019, arXiv e-prints, \href
  {https://ui.adsabs.harvard.edu/abs/2019arXiv190210045T} {p. arXiv:1902.10045}

\bibitem[\protect\citeauthoryear{{Torrence} \& {Compo}}{{Torrence} \&
  {Compo}}{1998}]{wavelet_theory}
{Torrence} C.,  {Compo} G.~P.,  1998, \mn@doi [Bulletin of the American
  Meteorological Society] {10.1175/1520-0477(1998)079<0061:APGTWA>2.0.CO;2},
  \href {https://ui.adsabs.harvard.edu/abs/1998BAMS...79...61T} {79, 61}

\bibitem[\protect\citeauthoryear{{Torres-Zafra}, {Cellone}, {Buzzoni},
  {Andruchow}  \& {Portilla}}{{Torres-Zafra} et~al.}{2018}]{torres-zafra2018}
{Torres-Zafra} J.,  {Cellone} S.~A.,  {Buzzoni} A.,  {Andruchow} I.,
  {Portilla} J.~G.,  2018, \mn@doi [\mnras] {10.1093/mnras/stx2561}, \href
  {https://ui.adsabs.harvard.edu/abs/2018MNRAS.474.3162T} {474, 3162}

\bibitem[\protect\citeauthoryear{{Urry} \& {Padovani}}{{Urry} \&
  {Padovani}}{1995}]{urry1995}
{Urry} C.~M.,  {Padovani} P.,  1995, \mn@doi [\pasp] {10.1086/133630}, \href
  {https://ui.adsabs.harvard.edu/abs/1995PASP..107..803U} {107, 803}

\bibitem[\protect\citeauthoryear{{VanderPlas}}{{VanderPlas}}{2018}]{vanderplas}
{VanderPlas} J.~T.,  2018, \mn@doi [\apjs] {10.3847/1538-4365/aab766}, \href
  {https://ui.adsabs.harvard.edu/abs/2018ApJS..236...16V} {236, 16}

\bibitem[\protect\citeauthoryear{{Vaughan}}{{Vaughan}}{2005}]{LSsignificance}
{Vaughan} S.,  2005, \mn@doi [\aap] {10.1051/0004-6361:20041453}, \href
  {https://ui.adsabs.harvard.edu/abs/2005A&A...431..391V} {431, 391}

\bibitem[\protect\citeauthoryear{{Villata} \& {Raiteri}}{{Villata} \&
  {Raiteri}}{1999}]{Villata99}
{Villata} M.,  {Raiteri} C.~M.,  1999, \aap, \href
  {https://ui.adsabs.harvard.edu/abs/1999A&A...347...30V} {347, 30}

\bibitem[\protect\citeauthoryear{{Villata} et~al.,}{{Villata}
  et~al.}{1997}]{villata1997}
{Villata} M.,  et~al., 1997, \mn@doi [\aaps] {10.1051/aas:1997313}, \href
  {https://ui.adsabs.harvard.edu/abs/1997A&AS..121..119V} {121, 119}

\bibitem[\protect\citeauthoryear{{Villata} et~al.,}{{Villata}
  et~al.}{2002}]{webt}
{Villata} M.,  et~al., 2002, \memsai, \href
  {https://ui.adsabs.harvard.edu/abs/2002MmSAI..73.1191V} {73, 1191}

\bibitem[\protect\citeauthoryear{{Villata} et~al.,}{{Villata}
  et~al.}{2009}]{villata2009}
{Villata} M.,  et~al., 2009, \mn@doi [\aap] {10.1051/0004-6361/200912732},
  \href {https://ui.adsabs.harvard.edu/abs/2009A&A...504L...9V} {504, L9}

\bibitem[\protect\citeauthoryear{{Vovk} \& {Neronov}}{{Vovk} \&
  {Neronov}}{2013}]{vovk2013}
{Vovk} I.,  {Neronov} A.,  2013, \mn@doi [\apj] {10.1088/0004-637X/767/2/103},
  \href {https://ui.adsabs.harvard.edu/abs/2013ApJ...767..103V} {767, 103}

\bibitem[\protect\citeauthoryear{{Wang}, {An}, {Baan}  \& {Lu}}{{Wang}
  et~al.}{2014}]{radio1156}
{Wang} J.-Y.,  {An} T.,  {Baan} W.~A.,   {Lu} X.-L.,  2014, \mn@doi [\mnras]
  {10.1093/mnras/stu1135}, \href
  {https://ui.adsabs.harvard.edu/abs/2014MNRAS.443...58W} {443, 58}

\bibitem[\protect\citeauthoryear{Witt \& Schumann}{Witt \&
  Schumann}{2005}]{wavelet}
Witt A.,  Schumann A.,  2005, Nonlinear Processes in Geophysics, 12, 345

\bibitem[\protect\citeauthoryear{{Zhang}, {Yan}, {Liao}  \& {Wang}}{{Zhang}
  et~al.}{2017a}]{zhang2017a}
{Zhang} P.-f.,  {Yan} D.-h.,  {Liao} N.-h.,   {Wang} J.-c.,  2017a, \mn@doi
  [\apj] {10.3847/1538-4357/835/2/260}, \href
  {https://ui.adsabs.harvard.edu/abs/2017ApJ...835..260Z} {835, 260}

\bibitem[\protect\citeauthoryear{{Zhang}, {Yan}, {Zhou}, {Fan}, {Wang}  \&
  {Zhang}}{{Zhang} et~al.}{2017b}]{zhang2017b}
{Zhang} P.-F.,  {Yan} D.-H.,  {Zhou} J.-N.,  {Fan} Y.-Z.,  {Wang} J.-C.,
  {Zhang} L.,  2017b, \mn@doi [\apj] {10.3847/1538-4357/aa7ecd}, \href
  {https://ui.adsabs.harvard.edu/abs/2017ApJ...845...82Z} {845, 82}

\bibitem[\protect\citeauthoryear{{de Diego}, {Kidger}, {Gonzalez-Perez}  \&
  {Lehto}}{{de Diego} et~al.}{1997}]{dediego1997}
{de Diego} J.~A.,  {Kidger} M.~R.,  {Gonzalez-Perez} J.~N.,   {Lehto} H.~J.,
  1997, \aap, \href {https://ui.adsabs.harvard.edu/abs/1997A&A...318..331D}
  {318, 331}

\makeatother
\end{thebibliography}

\appendix

\section{Optical V band analysis for 3C 66A}\label{appen}

\begin{figure*}
\centering
\subfigure{\includegraphics[width=0.85\columnwidth]{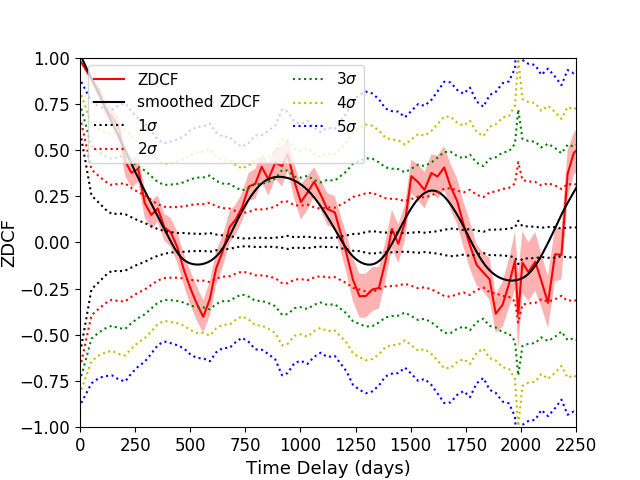}}
\subfigure{\includegraphics[width=0.85\columnwidth]{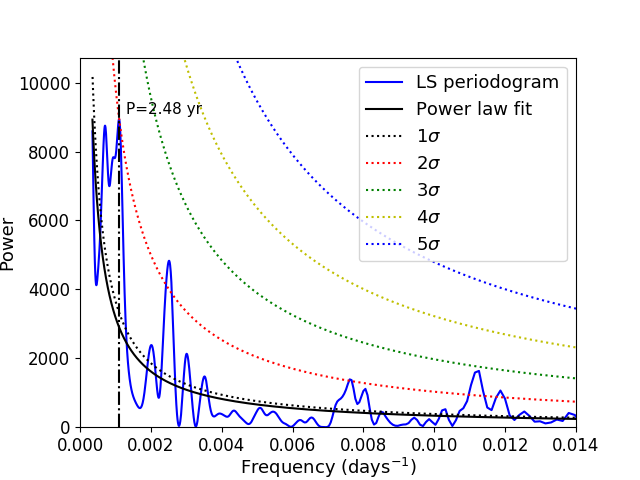}}
\subfigure{\includegraphics[width=0.85\textwidth]{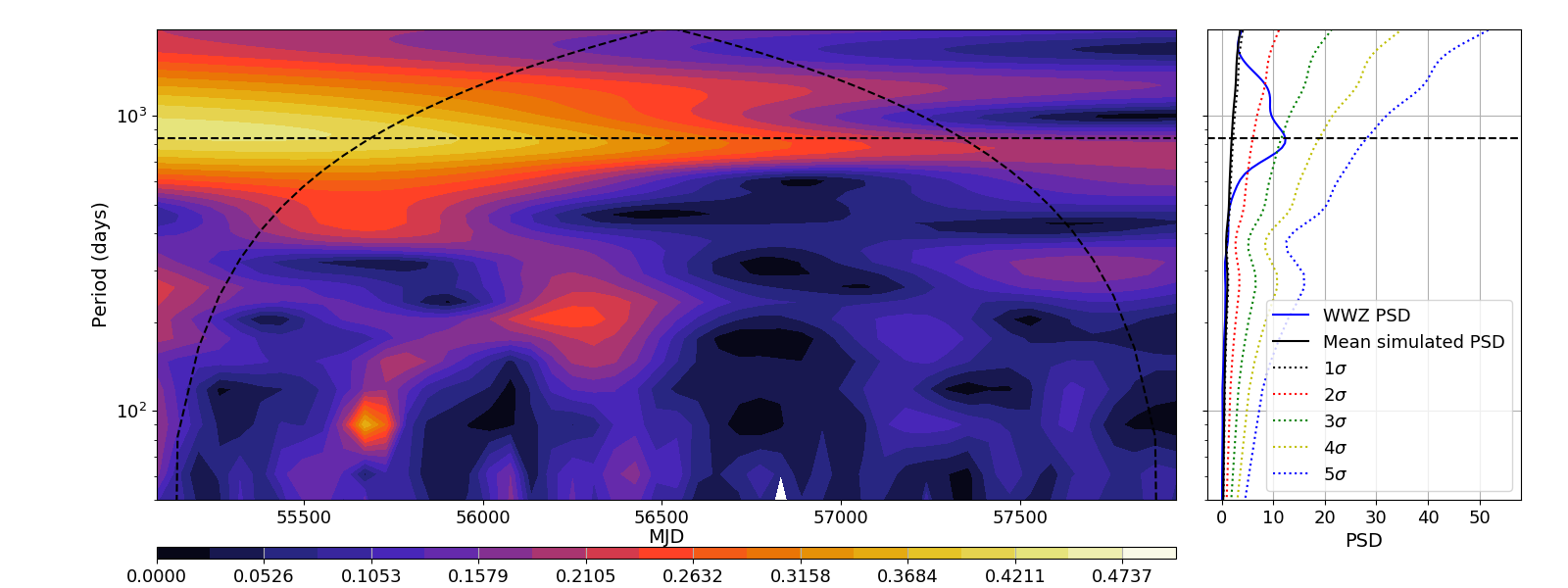}}
\caption{Periodicity analysis of the V band data of 3C 66A. \textit{Top left}: ZDCF method. The autocorrelation is given in red and the smoothed curve in black. \textit{Top right}: Lomb-Scargle periodogram. The periodogram is given in blue and the power law fit in black. \textit{Bottom}: WWZ diagram. The left panel shows the 2D power spectrum as a function of period and time. The black dashed curve represents the COI. The right panel shows the PSD in blue and the mean simulated PSD in black. The coloured dotted lines represent the different significance levels. The black horizontal dashed line marks the peak of the PSD.}
\label{3c66a_all_plots_V}
\end{figure*}

The search for periodicity in 3C 66A was also carried out in the V band in addition to the R band presented in Sections \ref{3cR} and \ref{3cpolR}. Fig. \ref{3c66a_all_plots_V} shows the results for the total optical magnitude. The ZDCF exhibits a periodic structure with maxima/minima reaching a CL of 2$\sigma$-3$\sigma$. The period derived from the ZDCF is 2.04$\pm$0.10 years. The LS periodogram presents a peak at a CL of $\sim$2$\sigma$ with a period of 2.48$\pm$0.29 years. The available data in V band cover only the low state period of the R band light curve, so this result is comparable to the R-band result for this period. Peaks with lower periods at a CL of $\sim$2$\sigma$ appear due to the gaps in the coverage and uneven sampling of the light curve. Finally, a peak is identified in the WWZ diagram with a period of 2.28$\pm$0.47 years and at a significance of $\sim$3$\sigma$.

\bsp	
\label{lastpage}
\end{document}